\documentclass[prd,aps,twocolumn,a4paper,floatfix,showpacs,nofootinbib]{revtex4-1}

\pdfoutput=1
\usepackage{graphicx,psfrag}
\usepackage[printwatermark]{xwatermark}
\usepackage{xcolor}
\usepackage{mathrsfs}
\usepackage{amsmath,amsfonts,amssymb,pifont,bm}
\usepackage{multirow,enumerate}
\usepackage{comment,hyperref}
\usepackage{color}
\usepackage{acronym}
\usepackage{xspace}
\usepackage[normalem]{ulem}
\usepackage{xfrac}
\usepackage[group-digits=false]{siunitx}
\usepackage{booktabs,makecell,multirow}
\usepackage{txfonts}
\usepackage{soul}
\newacro{BH}{black hole}
\newacro{NS}{neutron star}
\newacro{PN}{Post-Newtonian}
\newacro{BBH}{binary-black-hole}
\newacro{BNS}{binary neutron star}
\newacro{NSBH}{neutron-star black-hole}
\newacro{EOB}{effective-one-body}
\newacro{NR}{numerical relativity}
\newacro{GW}{gravitational-wave}
\newacro{EB}{enriched basis}
\newacro{SVD}{singular value decomposition}
\newacro{PSD}{power spectral density}
\newacro{aLIGO}{Advanced Laser interferometer Gravitational-wave Observatory}
\newacro{AZDHP}{aLIGO zero detuned high power density}
\newacro{GR}{general relativity}
\newacro{PE}{parameter estimation}
\newacro{LAL}{LIGO algorithm library}
\newacro{ROM}{reduced-order model}

\newcommand{\be}{\begin{equation}}
\newcommand{\ee}{\end{equation}}
\newcommand{\bea}{\begin{eqnarray}}
\newcommand{\eea}{\end{eqnarray}}
\newcommand{\bel}{\begin{align}}
\newcommand{\eel}{\end{align}}

\def\Msun{{\rm M_{\odot}}}
\def\GMc2{{\rm G M_{\odot} c^{-2}}}

\newcommand{\chieff}{\ensuremath{\chi_{\textrm{eff}}}}
\newcommand{\flow}{\ensuremath{f_{\textrm{low}}}}
\newcommand{\fhigh}{\ensuremath{f_{\textrm{high}}}}

\usepackage{color}
\definecolor{cyan}{rgb}{0,0.9,0.9}
\definecolor{orange}{rgb}{0.9,0.5,0}
\definecolor{magenta}{rgb}{1,0,1}
\definecolor{purple}{rgb}{0.8,0.4,0.8}
\definecolor{gray}{rgb}{0.5,0.5,0.5}
\definecolor{mygreen}{rgb}{0.1,0.8,0.1}
\definecolor{darkblue}{rgb}{0.0,0.0,0.6}

\newcommand{\AEIHannover}{Max Planck  Institute for Gravitational Physics
(Albert Einstein Institute), Callinstr.~38, 30167 Hannover, Germany}
\newcommand{\UniHannover}{Leibniz Universit\"at Hannover, D-30167 Hannover, Germany}

\begin{document}

\title{Enhancing gravitational waveform models through dynamic calibration}

%main work done by
\author{Yoshinta \surname{Setyawati}$^{1,2}$}
\author{Frank \surname{Ohme}$^{1,2}$}

%supporters
\author{Sebastian \surname{Khan}$^{1,2}$}

\affiliation{${}^1$ \AEIHannover}
\affiliation{${}^2$ \UniHannover}

\date{\today}

\begin{abstract}

%   This paper is aimed to describe a new method to generate a quick and more accurate gravitational waveform using singular value decomposition of principal component analysis.
%   We arrange each component of waveform family in a matrix of frequency domain. We then project them onto the basis of less accurate waveform family, interpolate and project them back to produce \sk{an} enrich\sk{ed} \sk{waveform} basis\sk{.}
%   results heading (mandatory)
%   We found that the mismatch of enrich basis is at least ten times better than the mismatch of the approximate waveforms given the uniform interpolation grid.

Current gravitational-wave observations made by Advanced LIGO and Advanced Virgo use theoretical models that predict the signals generated by the coalescence of compact binaries.
Detections to date have been in regions of the parameter space where systematic modeling biases have been shown to be small. However, we must now prepare for a future with observations covering a wider range of binary configurations, and ever increasing detector sensitivities placing higher accuracy demands on theoretical models.
Strategies to model the inspiral, merger and ringdown of coalescing binaries are restricted in parameter space by the coverage of available numerical-relativity simulations, and when more numerical waveforms become available, substantial efforts to manually (re-)calibrate models are required.
The aim of this study is to overcome these limitations. We
explore a method to combine the information of two waveform models: an accurate, but computationally expensive \emph{target} model, and a fast but less accurate \emph{approximate} model. In an automatic process we systematically update the basis representation
of the approximate model using information from the target model.
% The \emph{target} model could be a more accurate but computationally expensive
% model and only be known at discrete points in parameter
% space that makes its use prohibitive in standard analyses.
The result of this process is a new model which we call the \emph{enriched basis}. This new model can be evaluated anywhere in the parameter space jointly covered by either the approximate or target model, and the enriched basis model is considerably more accurate in regions where the sparse target signals were available.
Here we show a proof-of-concept construction of signals from non-precessing, spinning black-hole binaries based on the phenomenological waveform family. We show that obvious shortcomings of the previous PhenomB being the approximate model in the region of unequal masses and unequal spins can be corrected by combining its basis with interpolated projection coefficients derived from the more recent and accurate PhenomD as the target model. Our success in building such a model constitutes an major step towards dynamically combining numerical relativity data and analytical waveform models in the computationally demanding analysis of LIGO and Virgo data.
% We also set the accuracy of the expected output by iterating this process until the expected accuracy is achieved.
% Here we show that we can build a new basis with median mismatch accuracy of less than $10^{-4}$.
% From the above result, we conclude that our method is efficient to dynamically\sk{I'm not really sure why it's dynamic} calibrate basis waveform with the expected parameter ranges and the desired accuracy.
% We conclude that our method can be used to efficiently improve
% existing analytic waveform models in a controlled, systematic manner
% by incorporating information from more accurate but computationally expensive
% models.
% \sk{A better ending needed I think, something with more of a bang.}
% \sk{I may have gone overboard with italicising of approximate and target.}
\end{abstract}

\maketitle

%===============================
%		INTRODUCTION
%===============================

\section{Introduction}
\label{sec:intro}

The dawn of the \ac{GW} era began with the first detection of a \ac{BBH} merger
on September 14, 2015 \cite{PhysRevLett.116.061102} by the \ac{aLIGO} \cite{advancedLIGO}.
More \acp{BBH} \cite{PhysRevLett.116.241103, PhysRevLett.118.221101, GW170608, PhysRevLett.119.141101} and one \ac{BNS}
merger on August 17, 2017 \cite{TheLIGOScientific:2017qsa} have been observed by
\ac{aLIGO} and Virgo \cite{advanceVirgo} during their first two observing
runs.

The search for \acp{GW} requires coincident signals in at least two
instruments. In order to uncover signals of astrophysical origin hidden behind the instruments' noise,
their data are filtered with a large number of waveform templates
\cite{PhysRevD.93.122003}. More than one hundred thousand templates of
coalescing compact binaries were employed in \ac{aLIGO} \ac{GW}
searches during each of the first two observing runs. An order of magnitude more modelled waveforms
are then used to estimate the source parameters and their uncertainties.
%Recent direct observations of \ac{GW}
More accurate and efficient follow ups of \ac{GW} detections and their parameters will be needed for the following \ac{aLIGO} observing runs.
This implies the need for waveform models covering a wide range of parameter space that can be generated quickly.
% Here we underline this issue and emphasize that waveform models play a crucial role in \ac{GW} detection and \ac{PE}.

The \ac{GW} signal emitted by coalescing binaries
depends upon many
different parameters that are often grouped into intrinsic and
extrinsic parameters.
Intrinsic parameters are astrophysical parameters of the binary. These are two mass parameters: the chirp mass ($\mathcal{M}_c$) and the symmetric mass ratio $\eta$; eccentricity; tidal parameters for neutron stars; and the spin components of the two objects ($\vec \chi_{1}, \vec \chi_{2}$) that are often represented by the dominant, effective spin parameter (\chieff) in the case of non-precessing binaries.
The exact definition of these parameters will be introduced in section \ref{sec:method}.

In this study we focus on non-precessing \acp{BBH} for which the spins are (anti-)aligned with the binary's orbital angular momentum. The dimensionless tidal parameters are set to zero. Eccentricity has also been neglected in all \ac{aLIGO} \ac{GW} searches that employ modelled templates so far, mainly because for most plausible astrophysical formation scenarios, the binary is expected to have circularized by the time its \ac{GW} signal enters the \ac{aLIGO} frequency range. However,
future waveform developments might include the eccentricity of the binary.

In addition to these properties, extrinsic parameters define the location and orientation of the source relative to the observer, such as the luminosity distance ($D_L$), inclination angle ($\iota$), sky position (RA, Dec), polarization angle ($\Psi$), time of coalescence ($t_c$), and phase of coalescence ($\phi_c$) \cite{PhysRevD.47.2198}. For non-precessing systems, modifications in these parameters simply shift the waveform in time, phase or amplitude, and they are much simpler to model than changes in intrinsic parameters.

In order to predict \ac{GW} signals from binaries, one needs to solve the Einstein equation in \ac{GR}.
%\ac{GW} waveform is a theoretical result of \ac{GR} as the existence of \ac{GW} is a consequence of Einstein's equation in \ac{GR}.
Analytical approximations have been established in form of \ac{PN} expansions. These are asymptotic expansions in a small parameter such as the ratio of the characteristic velocity of the binary to the speed of light \cite{PhysRevD.81.064004, Blanchet}.
By the nature of the approximation, \ac{PN} expansions become increasingly inaccurate as the two bodies move closer to each other and faster, entering the strong gravity regime. At this stage, \ac{NR} simulations provide the only viable approach to solve the Einstein equation \cite{baumgarte}.
%In general, \ac{NR} waveforms are known to have exceptional accuracy compared to other waveform models, but they are computationally extremely expensive \cite{PhysRevLett.111.241104, Ajith:2012az, PhysRevD.88.024040, hinder}.
In general, NR waveforms can in principle be very accurate and the accuracy can be tested through different types of convergence tests, 
but they are computationally extremely expensive  \cite{PhysRevLett.111.241104, Ajith:2012az, PhysRevD.88.024040, hinder}.
%Hence, the superior accuracy of NR simulations compared to the latest generations of waveform models should not be taken for granted. 

Hence, many efforts in the past focused on bridging \ac{PN} and \ac{NR} \cite{PhysRevD.77.044020, PhysRevD.78.104007}, leading to a variety of \ac{EOB} and phenomenological waveform models that are used in \ac{aLIGO}'s analyses.
\ac{EOB} is an analytical method proposed by Buonanno and Damour \cite{PhysRevD.59.084006, PhysRevD.79.124028, PhysRevD.77.024043, PhysRevD.81.084041, PhysRevD.84.124052, PhysRevD.87.084035, PhysRevD.86.024011, PhysRevD.89.061501} which substantially reformulates \ac{PN} results into a new description of the binary coalescence beyond the inspiral phase.
A different approach was developed to build phenomenological models (see \ref{subsec:waveform_models}) that essentially model coalescing binaries using analytical fits of \ac{PN}-\ac{NR} hybrids.

However, both approaches depend on a number of tunable parameters and fits whose optimal form and values are determined through complex procedures that typically require a fair amount of human input. Therefore, updated models that incorporate new \ac{NR} data and improved analytical descriptions typically take years to develop.

%this fit calculation is very \textit{human-dependent} meaning that it needs to be re-calculated with more \ac{NR} waveforms.

A different method to generate an accurate waveform model is based on sophisticated interpolation methods
to create a \emph{surrogate} model~\cite{PhysRevD.96.024058, PhysRevD.95.104023, PhysRevLett.115.121102} of \ac{NR} waveforms.
These surrogate models have a high accuracy to the original \ac{NR} waveforms, however,
they are limited to the parameter space covered by the original simulations.
Although boundaries are constantly being expanded in parameter space, this modeling strategy relies on large amounts of computational power.
At the time of writing this article, the latest precessing surrogate model~\cite{PhysRevD.96.024058} is limited in mass ratio and dimensionless spin magnitude to $q\leq2$ and $|\chi|\leq0.8$, respectively.

Here we explore a complementary method of constructing a waveform model that combines the information of an existing (computationally efficient) model with more accurate waveforms that are only available in a limited set of points in the parameter space.
A future application of our method would be a dynamical (i.e., fully automized) update of an analytical model with \ac{NR} waveforms to produce a new waveform model that can be evaluated continuously and has a better accuracy than the original model.

To develop our method, here we employ two analytic phenomenological models: PhenomB \cite{PhysRevLett.106.241101} being the \emph{approximate, less accurate} model and PhenomD \cite{PhysRevD.93.044006, PhysRevD.93.044007} being the \emph{target, more accurate} model.

We use \ac{SVD} to decompose the approximate model into an orthogonal basis and update the basis coefficients using information from the more accurate model.
Similar ideas of using \ac{SVD} to improve waveform models have been presented by Cannon et al \cite{PhysRevD.82.044025, PhysRevD.87.044008, PhysRevD.84.084003} and P\"urrer \cite{Purrer, Purrer:2015tud}. 

Cannon et al explore the use of reduced-order \ac{SVD} in time domain.
However, they only use one-dimensional interpolation in mass components and consider a restricted parameter space with no spin. We use a similar technique, but consider frequency-domain waveforms, and we extend the method to a much greater parameter space including spin.

P\"urrer discusses the use of \ac{SVD} to build computationally more efficient \acp{ROM} of existing spinning, non-precessing \ac{EOB} models.
\acp{ROM} are now a standard tool to reduce the time taken to generate a waveform,
but the resulting accuracy is that of the original model, or slightly less in challenging points of the parameter space.

% This paper is organized as follows. In section \ref{sec:method} we explain the phenomenological models, parameter ranges, reduced order, interpolation, mass scaling and iteration method used in this study.
% Section \ref{sec:result} presents our results and key findings for equal-spin and double-spin.
% Finally, we conclude and discuss future works in section \ref{sec:conclusion}.

Throughout this article geometric units are used by setting $G=c=1$.

%===============================
%		METHOD
%===============================

\section{Methodology}
\label{sec:method}

\subsection{Waveform models}
\label{subsec:waveform_models}

The constantly increasing sensitivity of \ac{GW} interferometers demands ever more accurate models. Updating and improving models is a major tasks entering the era of \ac{GW} astronomy, and we present a first end-to-end test of a fully automatic tuning that in future will use \ac{NR} simulations to improve analytical models. Here, however, we start with a proof-of-concept using two phenomenological waveform models.

The phenomenological family is a set of approximate waveform models, written as closed-form analytical expressions in the frequency domain \cite{Ajith:2007qp, Ajith:2007kx, PhysRevLett.106.241101, PhysRevD.82.064016, Hannam:2013oca, PhysRevD.93.044007}. These models have been calibrated to \ac{NR} waveforms that naturally cover a limited region of the intrinsic parameter space. However, the most recent models \cite{Hannam:2013oca, PhysRevD.93.044007} have been shown to be perfectly suited for current \ac{BBH} observations with mass ratios close to unity.

As explained in section \ref{sec:intro}, we use PhenomB as the \emph{approximate, less accurate model} that we aim to update with information from PhenomD as the \emph{accurate target model}. PhenomB~\cite{PhysRevLett.106.241101} was the first (anti)-aligned spin model of this family, released almost simultaneously with an alternative description of the same parameter space, called PhenomC~\cite{PhysRevD.82.064016}.
Both models were calibrated up to mass ratios of 4 and \ac{BH} spins up to 0.75. 
They have known shortcomings when extrapolating beyond the region of calibration, especially towards more extreme mass ratios.
PhenomD is the most recent and most sophisticated version of aligned-spin phenomenological models. It has been calibrated to 19 \ac{NR} waveforms from the SXS collaboration \cite{SXS} and the BAM code \cite{PhysRevD.77.024027, 0264-9381-25-10-105006} that span mass ratios from unity up to $18$ and dimensionless spin magnitudes up to 0.85 (0.98 for equal-mass systems) \cite{PhysRevD.93.044006, PhysRevD.93.044007}.
% An additional feature of PhenomD is its modularity. The connection of the inspiral part and the merger-ringdown part is ensured to be continuously differentiable in both phase and amplitude. Using this construction, it is straightforward to improve either part of \ac{GW} waveform model independently.

The intrinsic parameters of relevance in the non-precessing case are the mass ratio $q$, or equivalently the symmetric mass ratio $\eta$,
\begin{equation}
 q = \frac{m_1}{m_2} \geq 1, \qquad \eta = \frac{m_1 \; m_2}{(m_1 + m_2)^2},
\end{equation}
as well as the dimensionless spin projections along the orbital angular momentum $ \chi_{1z}, \chi_{2z}$ (non-vanishing spin components perpendicular to the orbital angular momentum cause precession effects that we leave for future work). For vacuum solutions of Einstein's Equation, the total mass $M = m_1 + m_2$ is a simple scaling factor.

We emphasize that the spin degrees of freedom in a binary are commonly reduced in phenomenological models to the observationally relevant dominant parameter combinations.
Following the analysis in \cite{PhysRevLett.106.241101,PhysRevD.82.064016} for aligned-spin binaries, the dominant spin effect in \ac{GW} phase can be expressed as the weighted combination of individual \ac{BH} spins,
\begin{equation}
 \chieff = \frac{m_1 \, \chi_{1z}+ m_2 \, \chi_{2z}}{m_1 + m_2}.
\end{equation}
Apart from an overall time and phase, PhenomB exclusively depends on $\chieff$ and $\eta$.
PhenomD uses $\chieff$ for the coefficients that were tuned to \ac{NR} simulations,
however, through the inspiral and the final state portion of PhenomD inherits two-spin dynamics. 
% from the \ac{PN} inspiral.
In section~\ref{sec:doublespin} we will apply our method to the 3D problem $(\eta, \chi_1, \chi_2)$
and express our results in terms of the symmetric $(\chieff)$ and anti-symmetric $(\chi_a)$ spin
parameters where $(\chi_a)$ is defined as
\begin{equation}
 \chi_a = \frac{\chi_{1z} - \chi_{2z}}{2}. \label{eq:chi_a}
\end{equation}

In the following sections, we present the details of how to update PhenomB with the more accurate PhenomD waveforms in frequency domain for a given range of $\eta$, $\chieff$ and scaled by the total mass $M$.
The end result of this computation is a new waveform model that
is closer to its target waveforms.
We call this new family as the \emph{\ac{EB}} waveforms.

\subsection{Parameter ranges}
\label{subsec:param_ranges}

This exploratory study is designed to test our method across a wide range in parameter space. Here, we
essentially consider the range in mass ratio and spins where PhenomD has been calibrated to \ac{NR} waveforms (see section \ref{subsec:waveform_models}),
\begin{align}
\eta & \in [0.05, 0.25], & & \chieff \in [-1, 1].
\end{align}
We stress that this region in the parameter space includes a considerable part where PhenomB has not been calibrated, e.g., mass ratios above $4$ ($\eta < 0.16$). What we are going to show is that despite the fact that the underlying approximate model does not accurately describe signals in certain regions, using accurate signals to update the approximate basis representation can entirely fix that problem.

In order to fully determine the signals for our test case, we fix the following additional parameters,
\begin{align}
 M &= 50 \Msun, & \flow &=30\,{\rm Hz}, \nonumber \\
\Delta f &=0.1 \,{\rm Hz}, &
M \fhigh &=0.2,
\end{align}
where $\flow$ and $\fhigh$ are the values of the lowest and the highest frequency we consider, respectively, and $\Delta f$ defines the numerical discretization of the signal.
M$\fhigh$ = 0.2 is chosen to be slightly higher than the signal with the largest ringdown frequency
in our dense grid\footnote{The system with the highest ringdown frequency will be the equal-mass,
maximally spinning case $(\chieff=1)$ which has dimensionless ringdown frequency of $\sim 0.13$.}.
For $M=50 \Msun$, $\fhigh$ corresponds to $812$ Hz.

\begin{figure}
\centering
\includegraphics[width=\hsize]{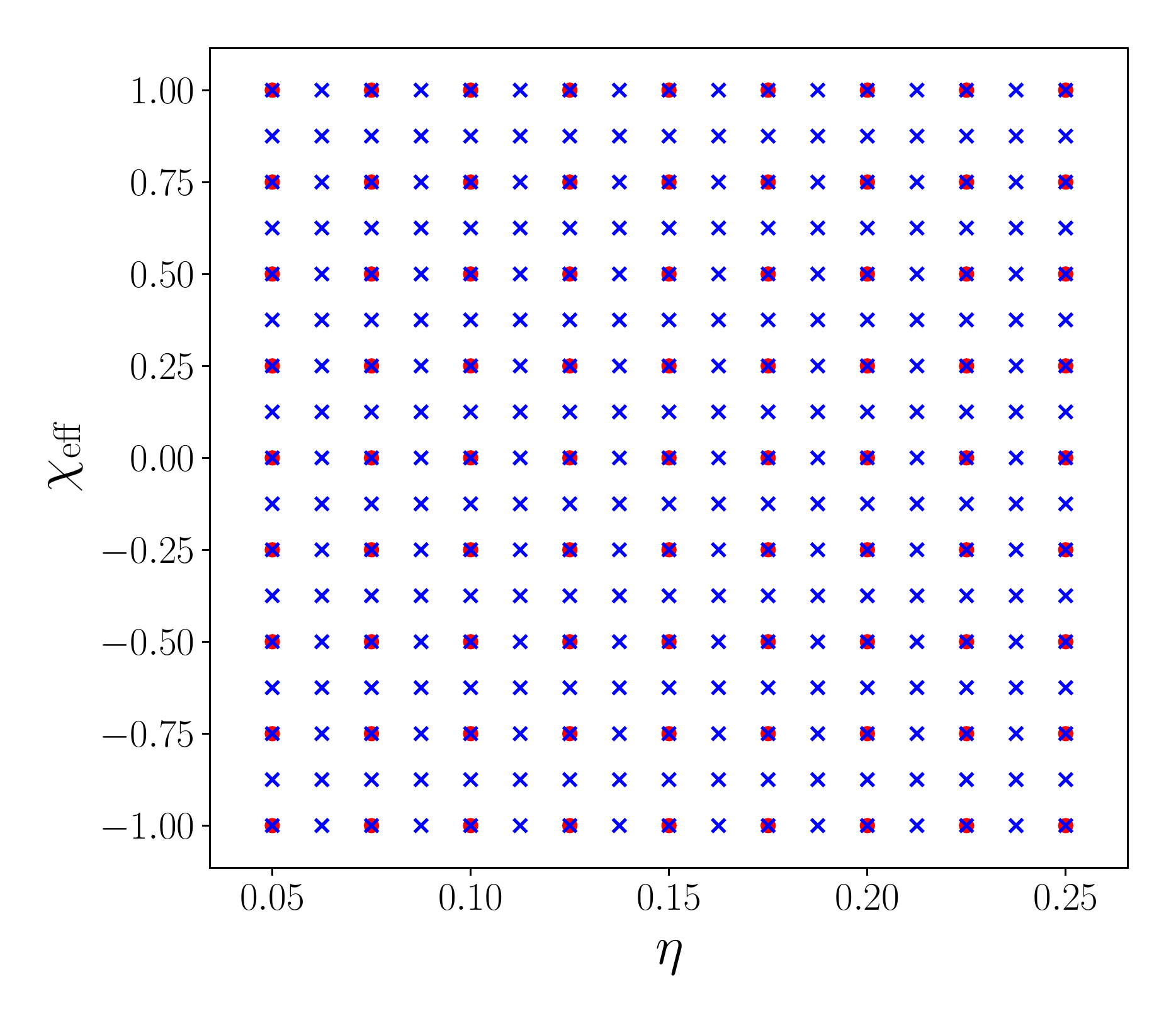}
\caption{Illustration of the two uniform grids we consider in $\eta$-$\chieff$ parameter space. 
The blue crosses illustrate the \emph{dense grid} of approximate signals that we use to build an \ac{SVD} basis, and the red circles are the \emph{sparse grid} of accurate signals we use to update the model.}
\label{grid_par}
\end{figure}

Following the above choice of parameter ranges, we create two two-dimensional (2D) uniform grids
in $\eta$ and $\chieff$.
We build a \emph{dense grid} of approximate PhenomB waveforms, and a \emph{sparse grid} of accurate PhenomD waveforms (see Fig.~\ref{grid_par} for visual representation).
Our dense grid contains $N = 65\times 65 = 4225$ signals, and the sparse
grid has $S = 33\times33 = 1089$ signals.
Thus, about 25\% of the approximate waveforms have the same $\eta$ and $\chieff$ as the target waveforms.

On each point of each grid, we generate the \ac{GW} polarizations, $h_{+/\times}$.
In this work, we only consider non-precessing signals and their $(\ell, |m|)=(2,2)$ multipoles which means that
the extrinsic parameters, such as the orientation and location of the binary, simply scale the amplitude of the signal and introduce a constant phase shift. We can treat these trivial dependencies independently and, at this stage, normalize all waveforms to have the same extrinsic parameters.
We then use the software library {\tt LALSuite} \cite{lalsuite} to generate the \ac{GW} polarizations.

In this study, both approximate and target models are
inexpensive to compute, so we can test our method for large numbers of
target-model waveforms.
In the future, we will use target waveforms that come from
computationally expensive methods such as \ac{NR} simulations. In that situation
we may not have access to signals at arbitrary points in parameter space
and we will have fewer waveforms.
Here we first choose a reasonably high number of target waveforms,
and later discuss how low this number can become to still produce satisfactory results.

%feedback
\subsection{Waveform matrices}
\label{subsec:wave_matrix}

In our method we will represent the waveform manifold of the \emph{approximate}
model with a set of
orthogonal basis functions computed using \ac{SVD}.
First, we prepare our dataset
in appropriate matrix form which we can factorize
subsequently.
The procedure is explained by the following steps.

\subsubsection{Waveform decomposition}

%\sk{I have simplified the layout and reduced the number of arguments in the equations.
%For example I don't think you need $(q,\chieff;M;Mf)$ every time. Also you generate at a fixed
%$M=50\Msun$ so why do you need $M$? Also you work with Hz right? So why $Mf$ and not just $f$.
%Sorry if it was my idea to be explicit before I don't agree with past Sebastian in that case.}

The frequency-domain strain $\tilde{h}(f)$ is the combination of both \ac{GW} polarizations , where $f$ is defined for positive frequencies. 
Here we assume the circular polarizations of \ac{GW} and describe $\tilde{h}(f)$ as follows:
\footnote{In other literature, the strain is sometimes defined as $h_+  - i h_\times$, owing to a different convention of the Fourier transform. Here we adopt the definition used in \ac{LAL}, $\tilde h(f) = \int h(t) \, \exp(-i 2\pi   f t) \, dt$}
\begin{equation}
\tilde{h}(f) = \tilde{h}_{+}(f) + i \, \tilde{h}_{\times}(f) \, .
\end{equation}

We note that if we express $\tilde{h}_{+/\times}(f)$ in terms of their amplitudes, $\mathcal{A}_{+,\times}$, and phases, $\Psi_{+,\times}$,
factoring out the dependency on the inclination angle, $\iota$, we obtain the following expressions \cite{PhysRevD.47.2198}:
\begin{eqnarray}
	\tilde{h}_{+}(f) &=& \mathcal{A}_{+}(f) \, e^{i \Psi_+(f)} \bigg(\frac{1+\cos^2 \iota}{2}\bigg) \,  ,  \label{eq:hplus} \\
	\tilde{h}_{\times}(f) &=& \mathcal{A}_{\times}(f) \, e^{i \Psi_\times(f)} \cos \iota \, .
\end{eqnarray}
The non-precessing signals we consider further satisfy a simple relation between the polarizations,
\begin{align}
 \mathcal{A}_{+} &= \mathcal{A}_{\times}, & \Psi_\times &= \Psi_+ - \frac{\pi}{2}. \label{eq:amp_phase_pol}
\end{align}
While (\ref{eq:amp_phase_pol}) is exactly valid only in the limit of large separations, assuming it through merger and ringdown is a commonly made approximation that does not introduce inaccuracies relevant to today's analyses.

By computing $\tilde h_+$ and $\tilde h_\times$ for $\iota = 0$, we can now decompose $\tilde h(f)$ into amplitude and phase components,
\begin{equation}
 \tilde h(f) = 2 \mathcal A_+(f) \; e^{i \Psi_+ (f)} ~.
\label{eq:strain_decomp}
\end{equation}
In this form, we can focus on two real-valued functions: the strain's amplitude
and phase (we drop the `+' subscript henceforth).
This decomposition is convenient because amplitude and phase are
simpler, real-valued, non-oscillatory functions which are better suited to perform \ac{SVD} than
the oscillating strain.

Once we have constructed the improved \ac{EB} amplitude $\mathcal{A}_{\rm EB}(f)$
and phase $\Psi_{\rm EB}(f)$, we can combine them again into the \ac{EB} strain
$\tilde{h}^{\rm EB} (f)$, as well as individual polarizations, using
Eqn.~(\ref{eq:hplus})-(\ref{eq:strain_decomp}).

\subsubsection{Phase alignment}

Time and phase shifts enter the frequency-domain waveform through the
\ac{GW} phase $\Psi(f)$ according to
\begin{equation}
	\Psi'(f) = \Psi(f) +  2 \pi f t + \psi,
\label{eq:timephaseshift}
\end{equation} %
where $t$ is the amount of time shifted and $\psi$ is the phase shift.

We use (\ref{eq:timephaseshift}) to align the phases in our approximate
waveform grid by determining the time and phase shift individually for each
configuration such that the square phase difference with one fiducial case is
minimized.
Specifically, we align the phases against the first case in our grid
($\eta=0.05, \chieff=-1$),
although any other choice yields comparable results.
By aligning the phases before performing the \ac{SVD} we remove variations
between the phases that are purely due to time and phase shifts. These
variations can always be re-introduced analytically via
(\ref{eq:timephaseshift}).
The shifted phase is denoted by $\Psi_B(f)$.

\subsubsection{Matrix form}

\begin{figure}
	\begin{minipage}{0.46\linewidth}
		\includegraphics[width=\linewidth]{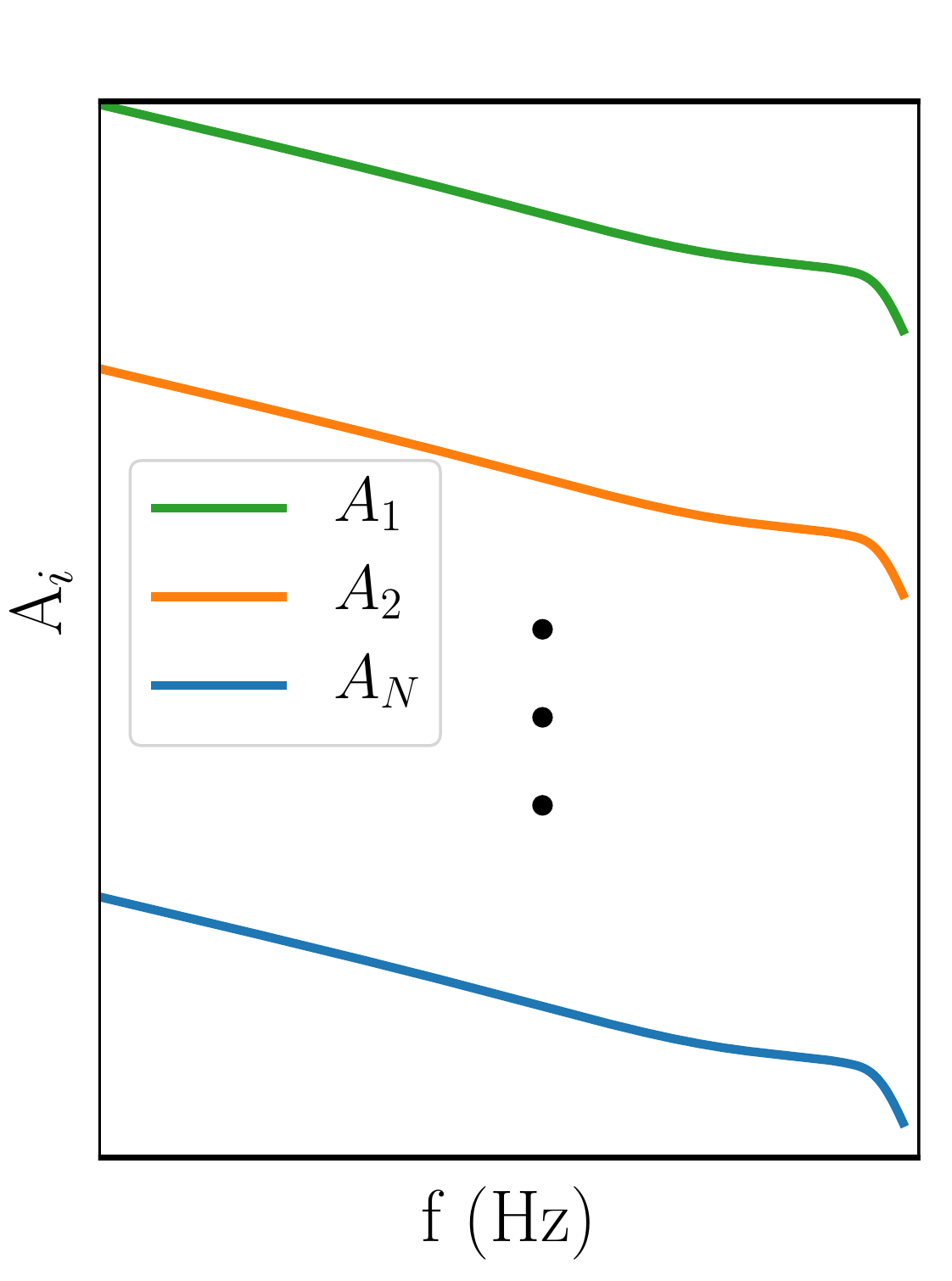}

	\end{minipage} \hfill
	\begin{minipage}{0.51\linewidth}
		\includegraphics[width=\linewidth]{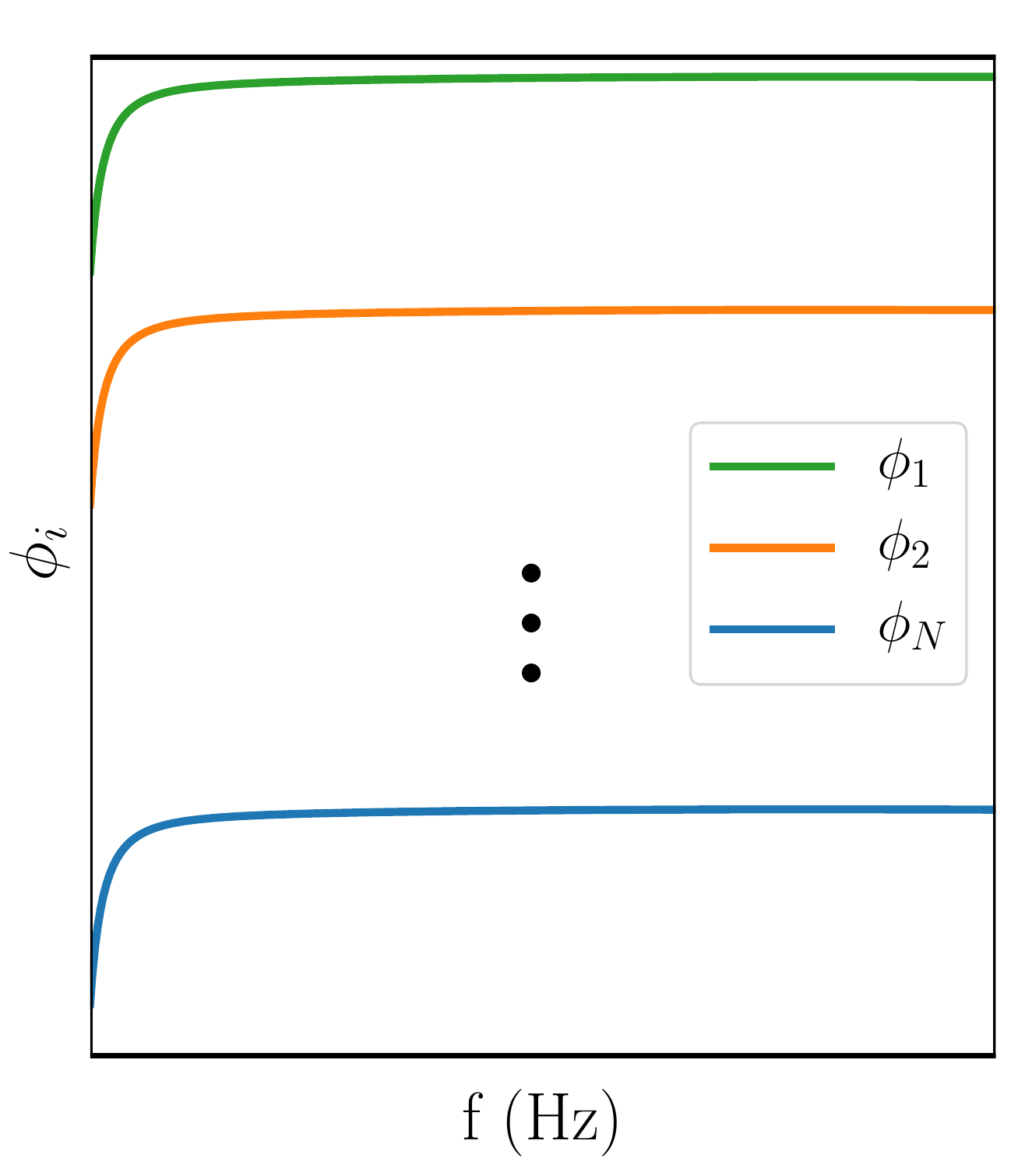}

	\end{minipage}
	\caption{Illustration of $N$ signals, each of length $L$, decomposed
into amplitudes (left) and phases (right) packed into two
matrices.}
\label{mtrx_ilustration}
\end{figure}

We generate $N$ signals from the approximate model PhenomB, each discretized at
$L > N$ points in frequency domain between $\flow$ and $\fhigh$. After
computing the strain $\tilde{h}(f)$ from the two polarizations as explained in
the previous section,
we decompose them into amplitude and phase then align the phases.
We then pack all amplitude and phase arrays into a matrix,
respectively (see Fig \ref{mtrx_ilustration})
Specifically, the rows of the matrices are arranged from the lowest ($\eta$,
$\chieff$) to the highest ($\eta$, $\chieff$).

We repeat the above procedure and generate $S$ target waveforms, using PhenomD,
on the sparse grid, where $S< N < L$.
At this point, we have four matrices: two amplitude matrices and two phase
matrices; one of each type for each approximant. The matrices of the
approximate model PhenomB have the dimensions $\mathbb{R}^{N \times L}$ while
the target PhenomD model matrices are $ \in \mathbb{R}^{S \times L}$.
With this prepared, we perform an \ac{SVD} as discussed in subsection
\ref{subsec:svd}.

\subsection{The Singular Value Decomposition}
\label{subsec:svd}

Our goal is to generate a new waveform family that can be evaluated for
arbitrary parameters from interpolating a set of sparse target waveforms.
To do this, we project our target model onto a basis of the approximate model,
generated from a grid that is as dense as
possible and computationally feasible.
As a first step, the basis is built by an appropriate factorization of the grid
of the approximate waveforms.

There are two main strategies to factorize sets of waveforms.
One uses a Gram-Schmidt orthogonalization to obtain the basis from a
first set of approximate waveforms
followed by a greedy algorithm to extend the basis until an acceptable
error limit is reached~\cite{PhysRevD.96.024058, PhysRevLett.106.221102, PhysRevD.94.044031}.
The second strategy uses the \ac{SVD} as in Cannon et al
\cite{PhysRevD.82.044025, PhysRevD.84.084003, PhysRevD.87.044008} and P\"urrer \cite {Purrer, Purrer:2015tud, PhysRevD.93.064041} to factorize each matrix into two unitary matrices
and one diagonal matrix with elements sorted in descending order.
The comparison between the two strategies has been discussed in
\cite{PhysRevD.96.024058}.
Here we use the \ac{SVD} because it produces smoother result, and
because it is elegant and convenient given that it sorts the contribution from
the dominant basis vector to the least important ones.
This ensures that the error caused by \ac{SVD} truncation is generally small.
% ($\leq$ $10^{-14}$).

%\ac{SVD} is widely known as a powerful method of matrix factorization and given in standard texts \cite{guide_book}.
%Various fields have applied this method for memory reduction and statistic.
%We will use \ac{SVD} to factorize the \emph{approximate} waveforms obtaining
%a basis representation of the approximate model.
%This strategy has been discussed in Cannon et al \cite{PhysRevD.84.084003, PhysRevD.87.044008} as mentioned in section \ref{sec:intro}.
%We discuss the projection coefficients and reduced basis in subsection \ref{subsec:projection_coef}, whereas the interpolation method is discussed further in \ref{subsec:interp}.

We adopt \ac{SVD} to individually factorize amplitude and phase matrices
($\bm{P}$) of PhenomB into two unitary
matrices ($\bm U$ and $\bm V$) and one diagonal matrix {$\bm \Sigma$} \cite{galassi},
\begin{equation}
\bm P = \bm U \bm \Sigma {\bm V}^T.
\label{singular_val}
\end{equation}
Here, $\bm U=[u_1| \ldots | u_p] \in \mathbb{R}^{N \times p}$ and
$\bm V = [v_1| \ldots | v_p] \in \mathbb{R}^{L \times p}$ are orthogonal
matrices and the superscript $T$ denotes the transpose of the corresponding
matrix.
The vectors $u_i$ and and $v_i$ are left and right singular vectors of
$\bm P$ respectively.
The singular values $\bm \Sigma=\operatorname{diag}(\sigma_1,\ldots,\sigma_p)
\in \mathbb{R}^{p \times p}$ is a diagonal matrix sorted in
descending order, where $p=\min (N, L)$, which in our setup
yields $p = N$. The diagonal elements
$\sigma_i^2$ are the eigenvalues of ${\bm P}^T \bm P$.

\ac{SVD} can be interpreted as matrix decomposition into a weighted sum of
separable matrices, meaning that a matrix $\bm P$ can be written as an
outer product of two vectors $\bm P= \bar{\bm u} \otimes \bm v^T$ ($\bar{\bm u}$
denote the $u$ vectors weighted by the singular values).
The rank of this outer product depends on how many singular values are involved
in the sum.
The index notation of the above reads
\begin{equation}
P_{ij} = \sum_{k=1}^p u_{ik} \; \sigma_k \; v_{kj}^T.
\label{eq:SVD_index_notation}
\end{equation}

\subsection{Projection coefficients and reduced order}
\label{subsec:projection_coef}

In our study, we use Eq.~(\ref{eq:SVD_index_notation}) in the following way.
Every row of the matrix $P_{ij}$ represents a Fourier-domain
series of either amplitude or phase; the index $j$ represents individual
frequency samples. Every one of those Fourier-domain series is expressed on
the right-hand side as a linear combination of orthogonal basis vectors
$(V^T)_{kj}$ ($k$ is the index of the basis, $j$ specifies the frequency)
multiplied with coefficients $c_{ik} =  u_{ik} \; \sigma_k$ ($k$ corresponds
to the associated basis, $i$ specifies the frequency series that is
reconstructed in this way). We call $c_{ik}$ the projection coefficients. The
projection coefficients can be interpreted as updating the left singular
vectors $u_{ik}$ weighted by the rank of singular value $\sigma_k$.

In
order to build an analytical model that can be evaluated continuously across
the parameter space, the projection coefficients need to become functions that
interpolate in the parameter space between the discrete points given in the
rows of $P_{ij}$. We emphasize
this below by replacing the index $i$ with the explicit functional dependency
on $\eta$ and $\chieff$, leading to
\begin{equation}
 c_k(\eta, \chieff) = \sum_{j=1}^L P_j(\eta, \chieff)\; V_{jk}.
\label{eq:proj_coefficients}
\end{equation}
The sum now describes a discretized inner product $\langle \cdot , \cdot
\rangle$, so that (\ref{eq:proj_coefficients}) becomes
\begin{equation}
 c_k(\eta, \chieff) = \left \langle P(\eta, \chieff),  v_{k} \right \rangle.
\end{equation}
Again, $P$ in this expression represents either the amplitude or phase for the
parameters $(\eta, \chieff)$, $v_k$ are the basis vectors calculated via
\ac{SVD}.

Different from standard practise in \ac{SVD} and \ac{ROM}, we now proceed by
calculating coefficients from projecting
the
\emph{target} waveforms' amplitude and phase onto the basis representation of
the \emph{approximate} waveform, respectively.
In addition, we study a reduction of the basis size that is achieved by only
considering the first $K$ coefficients. $K$ then reduces the rank of
the singular values matrices~\cite{PhysRevD.87.044008}, and it enters
(\ref{eq:SVD_index_notation}) as the upper limit of the sum instead of $p$.
This reduced order is introduced to increase computational efficiency and to
decrease memory requirements when building the \ac{EB} in comparison to the
full basis $k=N$.

By updating the approximate (less accurate) waveforms basis coefficients with
information from the (more accurate) target waveforms
we have manipulated the basis representation of approximate waveforms to be
closer to target waveforms.
Hence, we name this process \emph{enriching the basis}.

\subsection{Interpolation}
\label{subsec:interp}

%The \emph{enriched model} is only known at the points where the target waveforms were generated.
To construct our enriched basis model, we calculate the approximate \ac{SVD}
basis and project the target amplitude and phase onto the respective basis
vectors, giving us projection coefficients according to
(\ref{eq:proj_coefficients}) on the sparse grid in parameter space
(recall, the sparse grid is where we have access to accurate target signal). We
then interpolate the projection coefficients and calculate their values on all
points on the dense grid, so that we can compare with all approximate
signals that we needed to start this process.

We stress that the dimensionality of the interpolation depends on
the target model.
For equal-spin case, we use two-dimensional interpolation $(\eta,
\chieff)$ in parameter space, and later we consider two
independent spins, where we need three-dimensional interpolation.
Here we employ cubic spline interpolation as the most efficient and easy
method for this project.
However, different interpolation methods such as Chebyshev polynomials
\cite{PhysRevD.87.044008}, tensor product interpolation \cite{Purrer}, Gaussian
interpolation \cite{PhysRevD.96.123011} and empirical interpolation
\cite{PhysRevD.95.104023} have been used in different studies.
For the future, it will be beneficial to compare all these methods
systematically, evaluating computational efficiency, accuracy and
generalisability to higher dimensions.

Once the target waveform's coefficients, that we denote by $c'(\eta, \chieff)$,
have been obtained, we combine them with the basis vectors to
calculate the \ac{EB}'s amplitude and phase,
\begin{equation}
P_{j}^{\rm EB} (\eta,\chieff)= \sum_{k=1}^K c'_{k} (\eta,\chieff) v^T_{kj}.
\label{interp_EB}
\end{equation}
Having amplitude and phase, we can build $\tilde{h}^{\rm EB} (\eta ,\chieff)$
using Eq.~(\ref{eq:strain_decomp}).

\subsection{Match and improvement evaluation}
\label{subsec:matches}

Once the \ac{EB} strains $\tilde{h}^{\rm EB}$ have been calculated, we evaluate 
their
accuracy and improvement of \ac{EB} model relative to its approximate and 
target models.
We then test the accuracy of \ac{EB} model both at points where the
target model was used to update the projection coefficients,
as well as at points where no target signals were available and we use 
the interpolated projection coefficients.
To perform the evaluation, we compute \emph{matches} between PhenomB and
PhenomD and compare them to the matches between \ac{EB} and PhenomD.

The match is defined as the normalized, noise-weighted inner product between 
two waveforms $h_1$ and $h_2$~\cite{PhysRevD.47.2198},
maximised over relative time and phase shifts between them,
\begin{equation}
\mathcal{O} = \frac{\left \langle h_1,h_2 \right \rangle}{\|h_1\| \|h_2\|} = 
\max_{\phi_0,t_0} \left[4 \operatorname{Re} \int_{f_1}^{f_2} 
\frac{\tilde{h}_1(f) \, \tilde{h}_2^\ast (f)}{S_n(f)} \frac{df}{\|h_1\| 
\|h_2\|} \right] . \label{eq:overlap}
\end{equation}
Here, $\phi_0$ and $t_0$ are relative phase and time shifts between the 
waveforms, respectively, and $\|h\|^2=\langle h, h \rangle$. $S_n(f)$ is the 
noise spectral density of the detector, $\tilde h^\ast$ denotes the complex 
conjugation of $\tilde h$, and $(f_1,f_2)$ is a suitable integration range which corresponds to $\flow$ and $\fhigh$ respectively. 
%\fo{need to specify $f_1$ and $f_2$}
We use two noise spectra in our analysis, flat noise ($S_n \equiv 1$) and the
\ac{AZDHP} which is the anticipated design sensitivity of
\ac{aLIGO} in 2020 or later \cite{LIGO_design_sensitivity}.
The motivation behind using a flat \ac{PSD} is to evaluate the signal agreement 
with equal weight
on all frequencies independent of an assumed instrument, whereas using
\ac{AZDHP} allows us to relate our results to \ac{GW} analysis applications.

Matches are close to unity where waveforms agree (see section 
\ref{sec:result}), so it is easier to compare the difference between two models 
by quoting the mismatch, defined as
\begin{equation}
\mathcal{M} (h_1, h_2)=1-\mathcal{O}(h_1, h_2). \label{eq:mismatch}
\end{equation}
Finally, to quantify the accuracy improvement of the \ac{EB} model over the 
approximate model PhenomB, we define improvement, $\mathcal I$, as the 
mismatch of the approximate waveform with the target model divided by 
the mismatch of \ac{EB} model with the target, 
\begin{equation}
\mathcal{I}
(h_1, h_2)=\frac{\mathcal{M}(h_1, h_3)}{\mathcal{M}
(h_2, h_3)}, \label{eq:improvement}
\end{equation}
where in this study, $h_1, h_2$ and $h_3$ correspond 
to PhenomB, \ac{EB}, and PhenomD respectively.

% \subsection{Iteration}
% To construct a new family of waveform with a better accuracy than the approximate waveforms, we iterate the above process until the median mismatch reaches a desired limit $\mathcal{E}$.
% We then repeat the above methodology and compute the median mismatch and improvement.
% Subsequently, we project the target waveform to \ac{EB} of the last iteration and run it on and on.
% 
% This iterative procedure further refines the \ac{EB} model leading to
% a considerably more accurate model.
% We find that this process increases the accuracy of the original approximate waveform considerably.
% %Contrary, the time needed for iteration also grows as more accurate waveforms are being generated.
% %\sk{I'm tempted to remove this comment. If it's just because the interpolant is expansive I don't
% %think we need to mention this technical problem.}

%===============================
%		RESULT
%===============================

\section{Results}
\label{sec:result}

We have outlined a technique to build a more accurate waveform model in the 
above section.
Here we present results and analyses based on two different assumptions about 
the spins in the target parameter space, \textit{equal-spin} 
($\chieff$=$\chi_{1}$=$\chi_{2}$) and 
\textit{double-spin}, where $\chi_{1}$ and $\chi_{2}$ are varied independently 
(i.e., $\chi_a$ does not necessary vanish).

\subsection{Two dimensions: equal-spin systems}
\label{subsec:single_spin}

Following the above procedure, we evaluate the match between the \ac{EB} model 
and the target model under flat noise and \ac{AZDHP}.
We also compare the mismatch between the approximate model against the target 
model to calculate the improvement we gain.

\begin{figure}
\centering
\includegraphics[width=0.95\hsize]{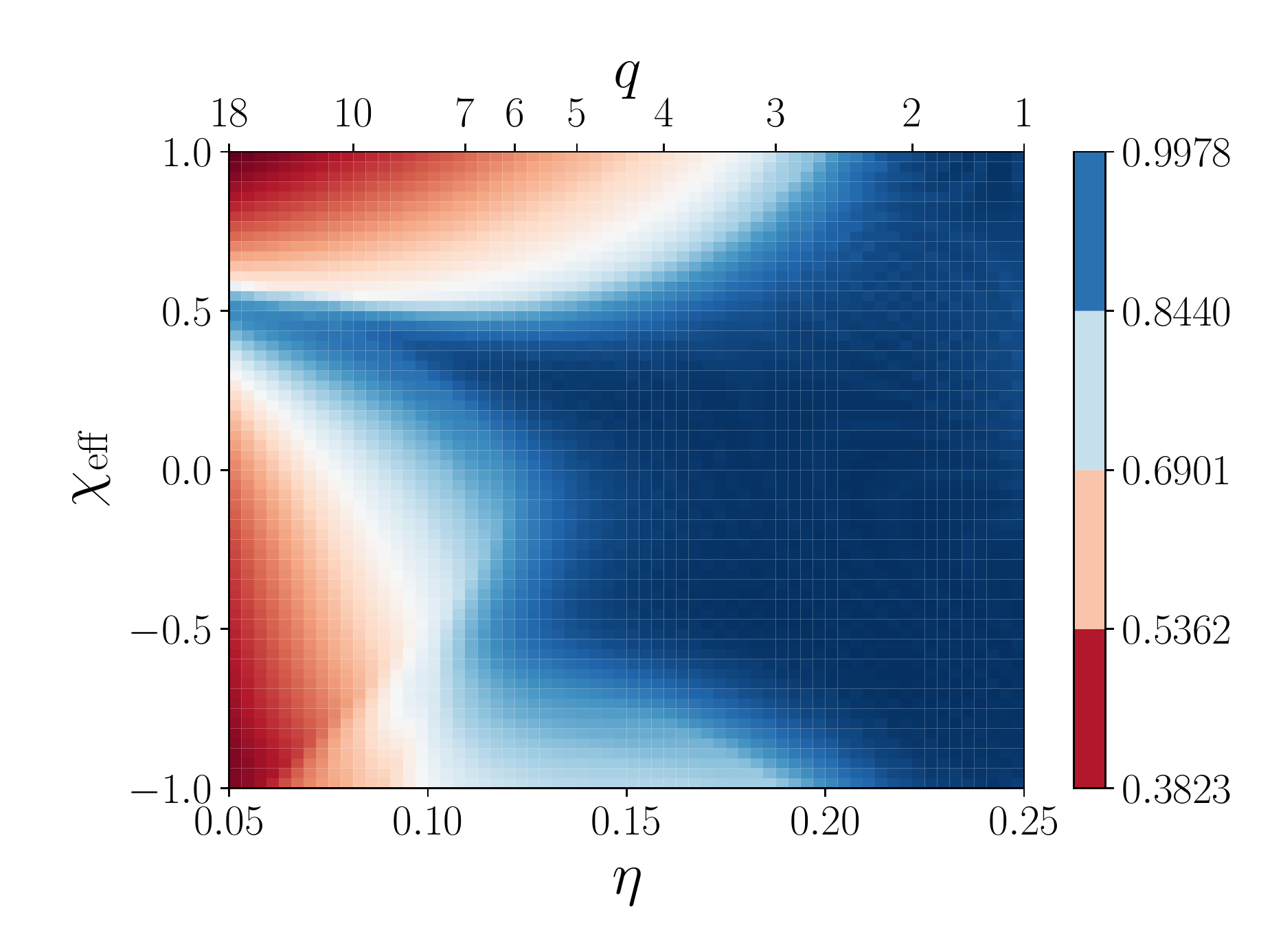}
\caption{Matches of PhenomB against PhenomD under flat \ac{PSD}.}
\label{pheB}
\end{figure}

Fig.~\ref{pheB} shows the original match of PhenomB against PhenomD. It is 
evident that PhenomB has not been calibrated to mass ratios above 4, and the 
agreement between the two models deteriorates quickly, especially for high 
spins.

\begin{figure}
\centering
\includegraphics[width=0.95\hsize]{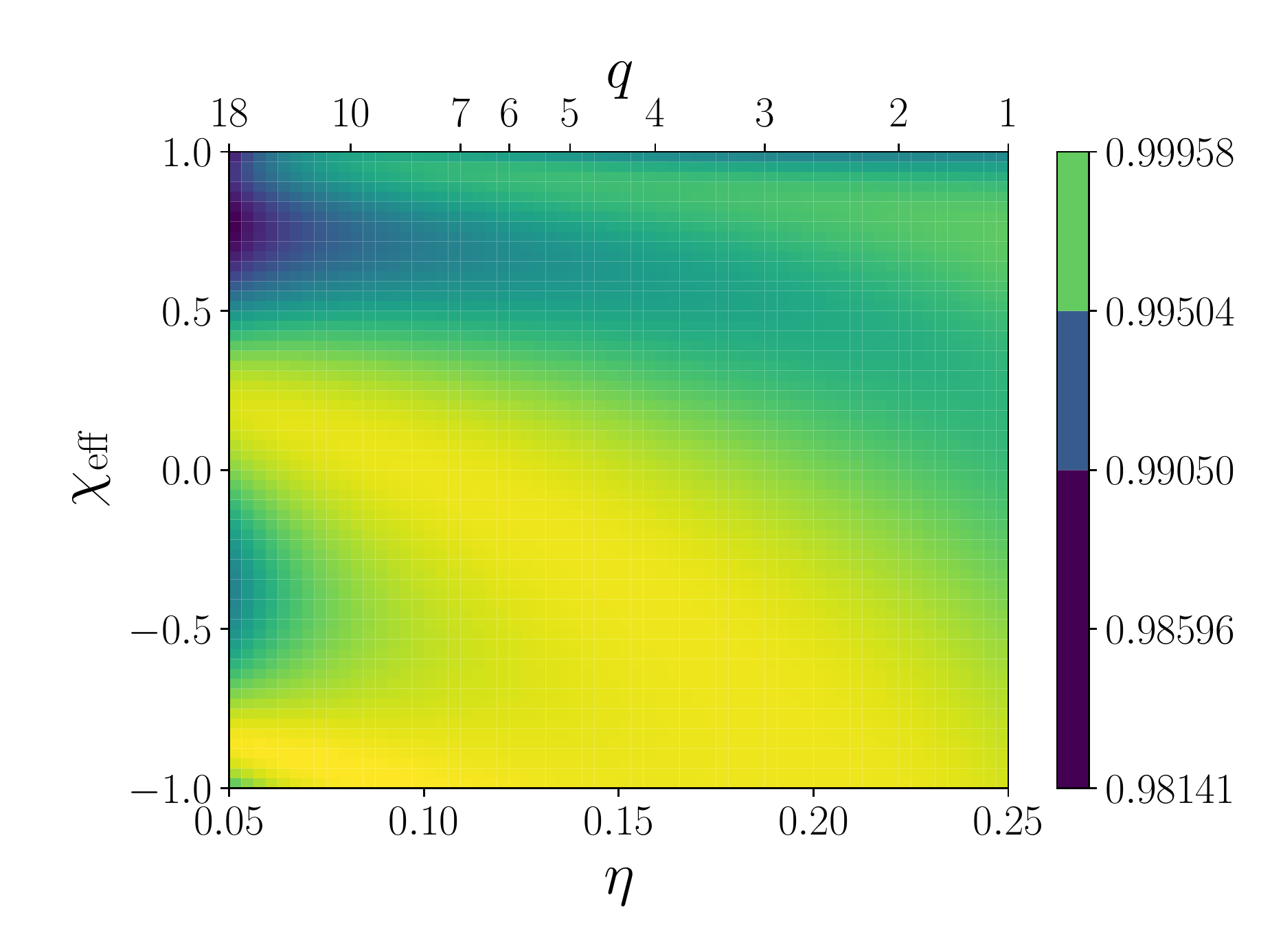}
\caption{Matches of \ac{EB} against PhenomD without interpolation and under flat 
\ac{PSD}.
In this figure, we generated target model in the same grid as the approximate 
model, and run our method in full bases (without reduced order).
This plot is used as comparison to interpolation and reduced order result as 
explained in the text.
}
\label{without_interp}
\end{figure}

Fig.~\ref{without_interp} presents the matches of \ac{EB} against PhenomD 
without invoking any interpolation. Recall, \ac{EB} here is based on basis 
vectors derived from PhenomB that do not accurately represent high-mass ratio
systems. However, by projecting $N = 65 \times 65$ PhenomD waveforms onto the 
basis derived from $N$ PhenomB signals on the same points in parameter space, 
we see that there is enough extra freedom in the basis such that updated 
projection coefficients can correct for the inaccuracies of the approximate 
model. Put differently, the space spanned by the approximate PhenomB 
basis vectors does contain more accurate signals, also for higher mass ratios, 
if the coefficients in front of the basis vectors are adapted appropriately. 
This might not be a surprising result, given the fairly large number of basis 
vectors we use; it is not a trivial result either.

\begin{figure}
\centering
\includegraphics[width=0.95\hsize]{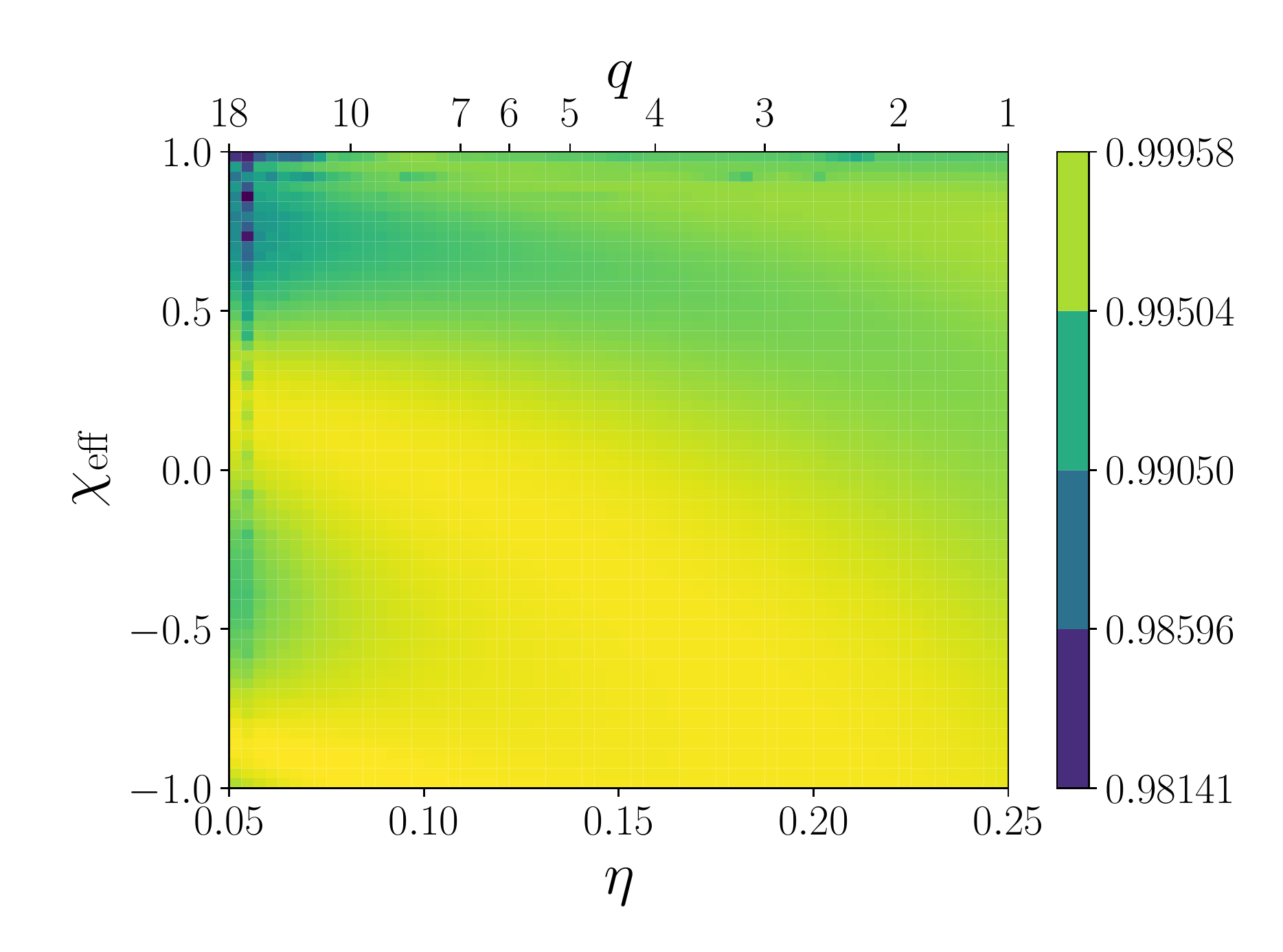}
\caption{Matches between \ac{EB} against PhenomD with interpolation and under flat 
\ac{PSD}.
In this figure, we generated both target and approximate models in regular grid .
The number of target model is about 25\% of the approximate model as explained in 
subsection \ref{subsec:param_ranges}.
We perform two dimensional interpolation (see subsection 
\ref{subsec:interp}) over the projection coefficients.
To make the comparison easier, we set the range of match equal as that on 
Fig.~\ref{without_interp}.
}
\label{interp}
\end{figure}

Of course, this is not a useful application of the method we develop. If one 
has access to $N$ accurate waveforms, there is no need build an approximate 
basis first. Now we reduce the number of accurate waveforms to $S \approx N/4$, and 
interpolate the projection coefficients to calculate \ac{EB} signals on all $N$ 
grid points. The mismatch result is shown in Fig.~\ref{interp}. In most parts 
of the parameter space, the accuracy of \ac{EB} is only very slightly lower 
than what was achieved in the ideal scenario shown in 
Fig.~\ref{without_interp}. Interpolation therefore does not introduce 
significant errors for the grids chosen here. Only at the boundaries of the 
parameter space we find higher mismatches in Fig.~\ref{interp}.

We note that interpolation will likely become a major source of error when the 
number of available target waveforms is decreased significantly and when the 
dimensionality of the parameter space increases. We shall return to 
discussing both issues later in this paper.

We have repeated the study with the \ac{AZDHP} noise curve and find 
qualitatively the same behaviour. A summary of mismatches (in $\log_{10}$ 
scale) and improvements are given in 
Table~\ref{tab_match_improve}.
We present the minimum, maximum and median mismatches across the dense grid, as 
well as the improvements defined by (\ref{eq:improvement}).

\begin{table}[t]
  \centering
  \caption{Mismatches between PhenomB and PhenomD as well as mismatches between 
\ac{EB} and PhenomD in $\log_{10}$ scale.
The improvement, $\mathcal I$, is defined by (\ref{eq:improvement}).
Here we compare the results using two different \acp{PSD}, flat \ac{PSD} 
($S_n=1$) and \ac{AZDHP}.
We also compare results that interpolate from the sparse to the 
dense grid with calculations entirely carried out on the dense grid (no 
interpolation).}
 \label{tab_match_improve}
\begin{ruledtabular}
\begin{small}
\begin{tabular}{llrrrrrr}
\multirow{2}{*}{PSD} & % \multirow{2}{*}{\specialcell{mismatch/ \\ 
% improvement}}
& \multicolumn{3}{c}{no interpolation} & 
\multicolumn{3}{c}{interpolation} \\
\cmidrule(l){4-8}
\multicolumn{1}{c}{}
 &
 \multicolumn{1}{c}{}
 &
    min & max & med &
    \phantom{XX}~min & max & med\\
\hline
\addlinespace[0.2cm]
\multirow{2}{*}{Flat}  & PhenomB  & -2.67   &   -0.001   &   -0.03  \\
\multicolumn{1}{c}{}   & \ac{EB} & -3.37   &   -1.99   &   -2.69   &   -3.38    
   & -1.73  &   -2.68   \\
\multicolumn{1}{c}{}    &  $\mathcal I$ &   1.42   &   1201   &   40   &   1.42 
       & 1195  &   39    \\

\addlinespace[0.3cm]
\multirow{2}{*}{\ac{AZDHP}} & PhenomB  &   -2.50    &   -0.10    &   -1.14 &      \\
\multicolumn{1}{c}{}  & \ac{EB} &   -3.23    &   -1.95    &   -2.71   &   -3.23  
 & -1.68  & -2.71    \\
\multicolumn{1}{c}{}  & $\mathcal I$    &  1.17        & 1082  &   39   &   
1.20   & 1082  & 39    
 \end{tabular}
 \end{small}
 \end{ruledtabular}
\end{table}

\begin{figure}
\centering
\includegraphics[width=0.95\hsize]{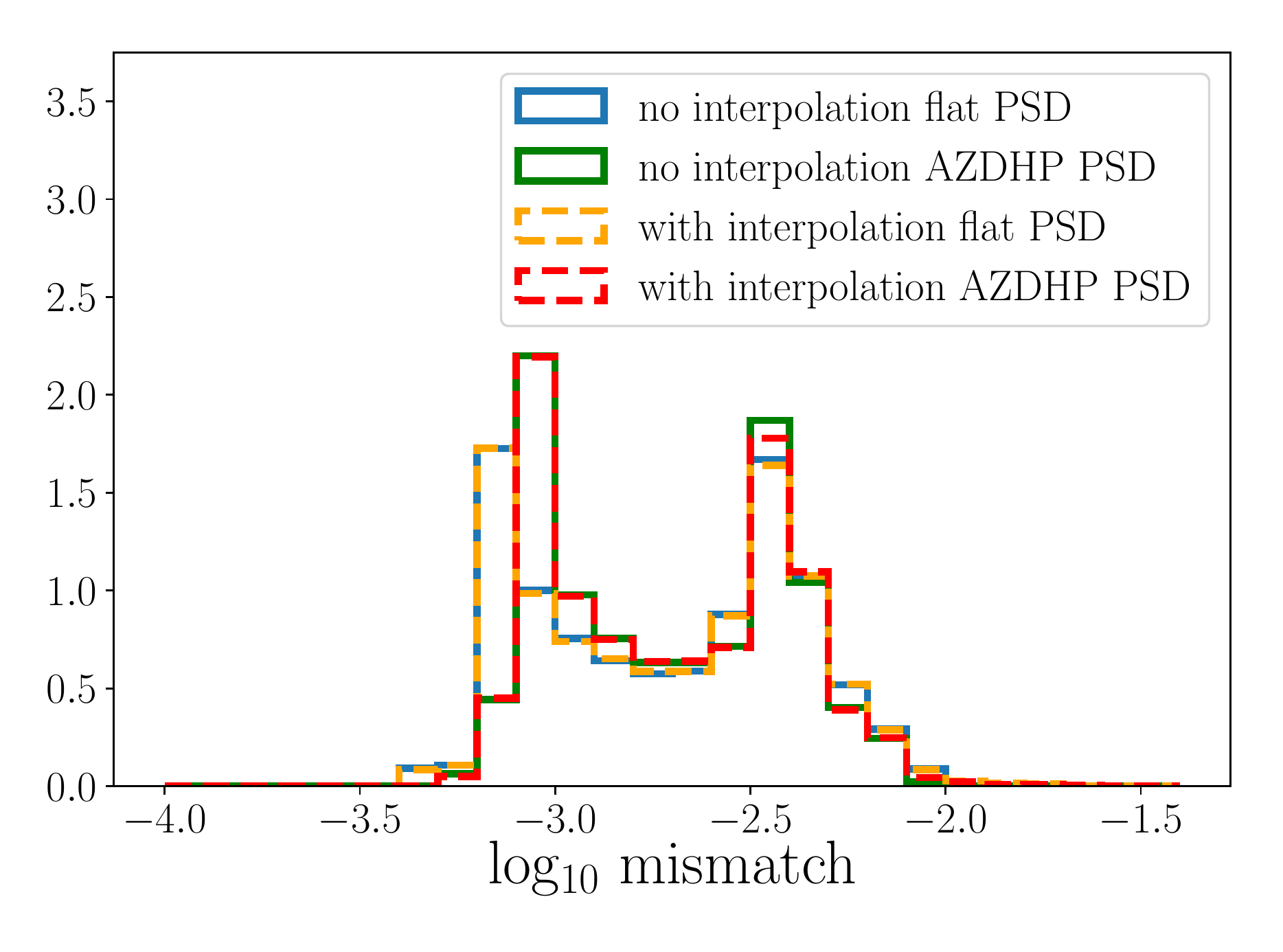}
\caption{Mismatches between \ac{EB} and PhenomD target signals for 
different configurations and \acp{PSD}. The histograms are normalized so 
that the sum of area under each line are set equal to unity. The dashed lines 
represent the result using fewer target signals and interpolation, whereas the 
respective solid lines show results using more target signals and no 
interpolation (see text).
}
\label{fig:match_histograms}
\end{figure}

Table~\ref{tab_match_improve} shows that overall the difference of mismatches 
using one or the other noise spectrum is relatively small. Full histograms are 
shown in Fig.~\ref{fig:match_histograms}. Because results are so similar, 
we only show the figures for the flat 
\ac{PSD}.

So far, we have generated our target model, PhenomD, on a regular grid in the 
parameter space as illustrated in Fig.~\ref{grid_par}.
We also investigate how the choice of positions of target signals affect 
our result.
For that reason, we distribute the same number of PhenomD waveforms 
randomly, drawn uniformly from the parameter space of $\eta$ and $\chieff$.
These target waveforms are then projected onto basis vectors coming from the 
dense regular grid of PhenomB signals.
We follow the above procedure to build the \ac{EB} coefficients  
and interpolate them onto the dense regular grid to evaluate mismatches between 
\ac{EB} and PhenomD waveforms with the same parameters.
Since the results for flat and \ac{AZDHP} \acp{PSD} are relatively close, we 
evaluate the mismatch assuming a flat \ac{PSD}. We find that $\log_{10} 
\mathcal M$ of random uniform grid ranges 
between $-1.39$ and $-3.39$.
For direct comparison, the mismatch of the regular grid of target waveforms is 
between $-1.73$ 
and $-3.38$ as presented in Table.~\ref{tab_match_improve}.
From this simple study, we argue that 
different positions will not affect the result significantly, so long as 
the number and distribution of parameters are similar.

\subsubsection{Accuracy of the reduced basis}
\label{subsubsec:reduce_faithfulness}

Here we examine the accuracy of \ac{EB} when restricting
ourselves to the first $K$ bases.
The advantage of a reduced basis is mainly to optimize computational power.

\begin{figure}
\centering
\includegraphics[width=\hsize]{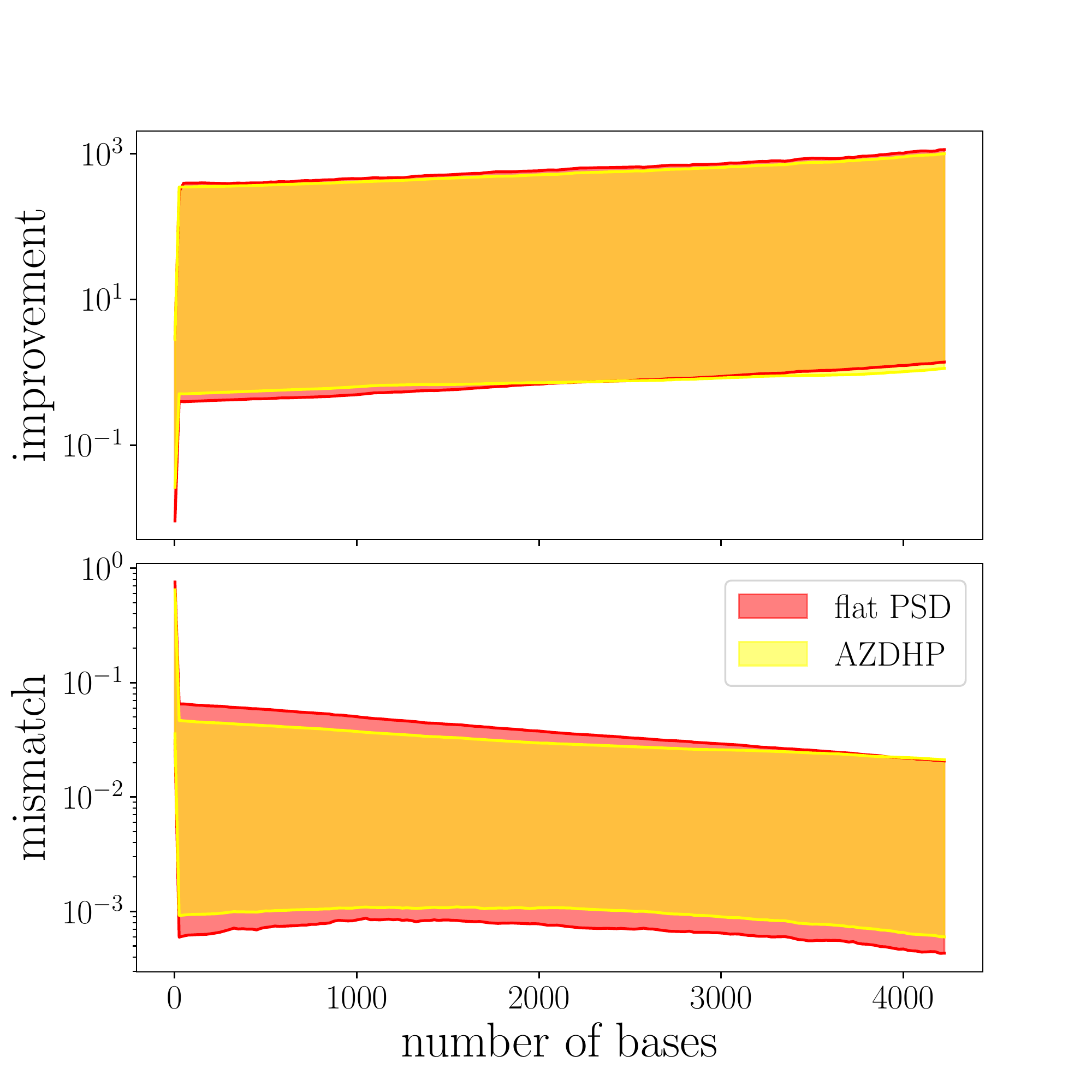}
\caption{The accuracy of a reduced-order model.
The top plot is the mismatch between \ac{EB} and PhenomD waveforms as a 
function of the number of reduced bases. The bottom plot shows the 
improvement, cf.\ (\ref{eq:improvement}).
The shaded areas are bounded by the minimum and maximum mismatches. 
The red area is obtained with a flat \ac{PSD} while the yellow area uses
\ac{AZDHP}. 
Results with different \acp{PSD} overlap well.
From this plot, using the minimum of 3375 bases, we can guarantee that all the 
\ac{EB} waveforms are more accurate than their approximate waveforms.
}
\label{red_basis_n_PSD}
\end{figure}

Fig.~\ref{red_basis_n_PSD} shows the mismatches and improvements as a 
function of the number of bases that are kept from the \ac{SVD} of PhenomB. 
To obtain the result, we projected $S \approx N/4$ PhenomD signals onto the 
PhenomB basis and performed interpolation 
as explained in previous section. For very small numbers of bases we observe a 
rapid drop in 
mismatches. After the first 25 
bases are included, however, the improvement of \ac{EB} 
is much more gradual when more bases are used. We speculate that the 
most important variations in PhenomB signals are already well described with 
25 basis vectors, but we do need a lot more basis vectors to accommodate 
additional features present in PhenomD that are not captured accurately by 
PhenomB (most notably, the high mass ratio, high spin regime).

If our goal is that the \ac{EB} signals are at least as accurate as the 
approximate model, and in most points of parameter space significantly more 
accurate, then we find that 3375 of 4225 bases are needed to guarantee 
that the improvement $\mathcal I \geq 1$. 

We might expect that higher parameter-space dimension ($\mathcal{D}+1$) require 
a larger number of bases to 
obtain at least the same mismatch as lower dimension ($\mathcal{D}$). 
Na\"ive intuition would be that the increase of dimensionality in parameter 
space requires an exponential growth of the basis size.
This is called ``curse of dimensionality``.
However, a study by Field et al 2012 \cite{PhysRevD.86.084046} shows that one
may only need a small number of additional bases in higher dimension to obtain 
comparable result as in lower dimension.
Therefore, the number of reduced bases is not exponentially proportional to the 
number of dimensions in parameter space.
Higher parameter-space dimensions, however, affect
computational time as we generate more waveforms covering a greater space.

\subsubsection{Minimum target waveforms}
\label{subsubsec:min_target}

In the analysis above, we used a uniform grid for target and approximate 
waveforms with the ratio of PhenomD to PhenomB signals of about $1/4$.
In this section, we explore the minimum number of target waveforms needed to 
obtain a computational efficient \ac{EB} model that improves the approximate 
model significantly.

\begin{figure}
\centering
\includegraphics[width=\hsize]{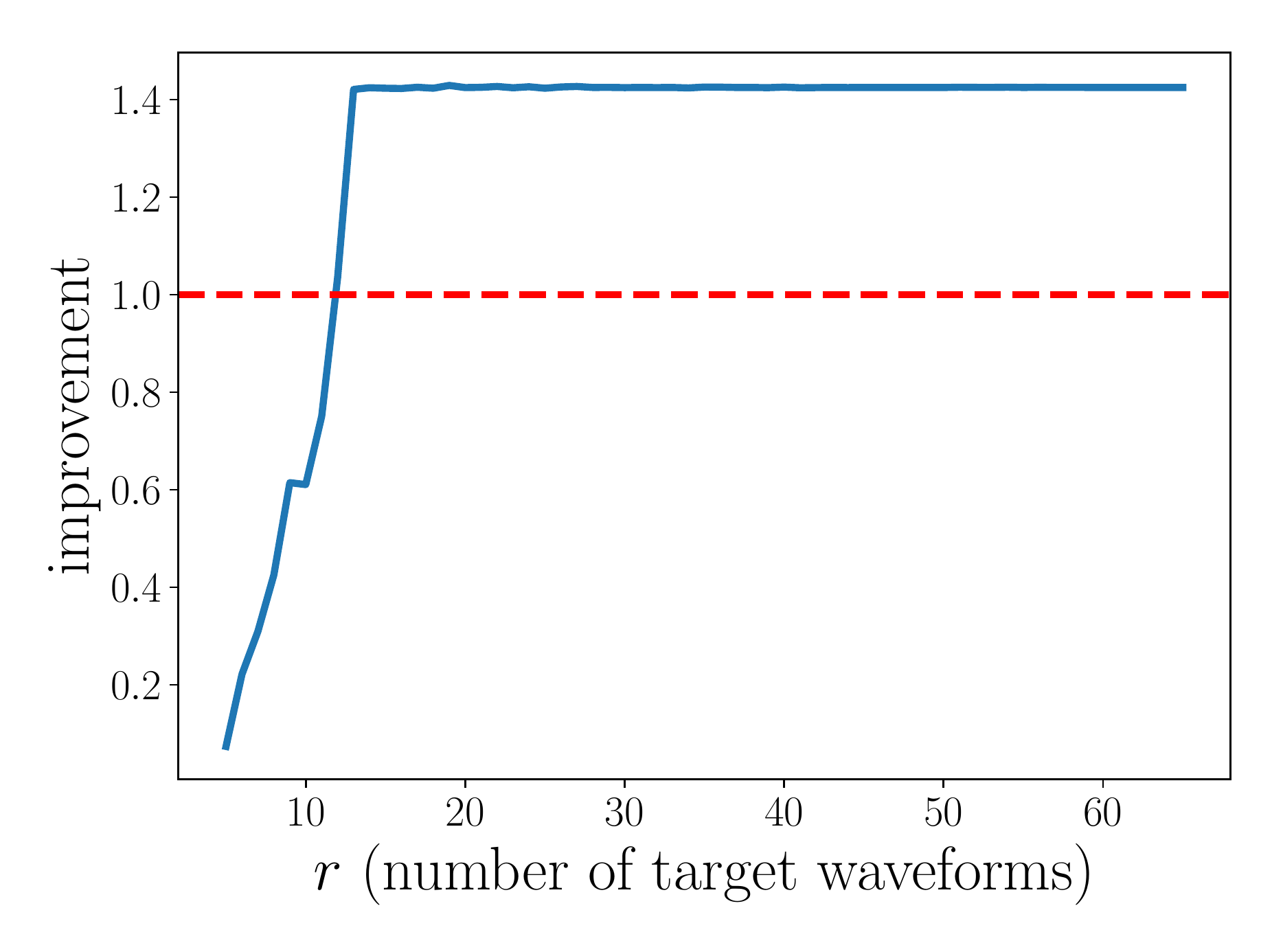}
\caption{Only 144 (12 $\times$ 12) PhenomD signals on a uniform grid are needed 
to guarantee that all \ac{EB} waveforms perform better than PhenomB (assuming 
flat 
noise).
The blue line is the value of the minimal improvement using $r \times r$ target 
waveforms. 
\label{fig:min_target}
}

\label{target_waves}
\end{figure}

We projected various numbers of target signals given on a sparse, uniform 
grid with $S = r \times r$ points onto the basis derived 
from the full $N$ approximate waveforms. We evaluate the improvement $\mathcal 
I$ on the dense grid (after interpolating the projection coefficients from the 
sparse onto the dense grid) and show the minimum in 
Fig.~\ref{fig:min_target}. We find that $12 \times 12 = 144$ target waveforms 
guarantee that all \ac{EB} results are better than PhenomB. This number is 
almost 30 times smaller than the number of PhenomB signals we use, and more 
than 95\% of the signals generated on the dense grid to compute mismatches are 
now interpolated and have not been used as target waveforms in the construction 
of \ac{EB}.

In fact, we find that the \ac{EB} model built with $S = 144$ accurate PhenomD 
signals performs in large parts of the parameter space comparable to the 
previous case of $33 \times 33$ target signals. Only the problematic boundary 
regions that were visible already in Fig.~\ref{interp} become more pronounced, 
both in size and mismatch. Better results, even with this relatively small 
number of target signals, can be achieved by the iteration procedure we will 
introduce below.

\subsubsection{Phase and amplitude contributions}
\label{subsubsec:phase_amp}

In order to identify the dominant contribution to the inaccuracies that we
reported for our \ac{EB} model, we now evaluate mismatches for individual  
components. In particular, we can apply the definition of the overlap 
(\ref{eq:overlap}) and mismatch (\ref{eq:mismatch}) to the amplitude alone, 
without maximizing over time and phase shifts. 

\begin{figure}
\centering
\includegraphics[width=\hsize]{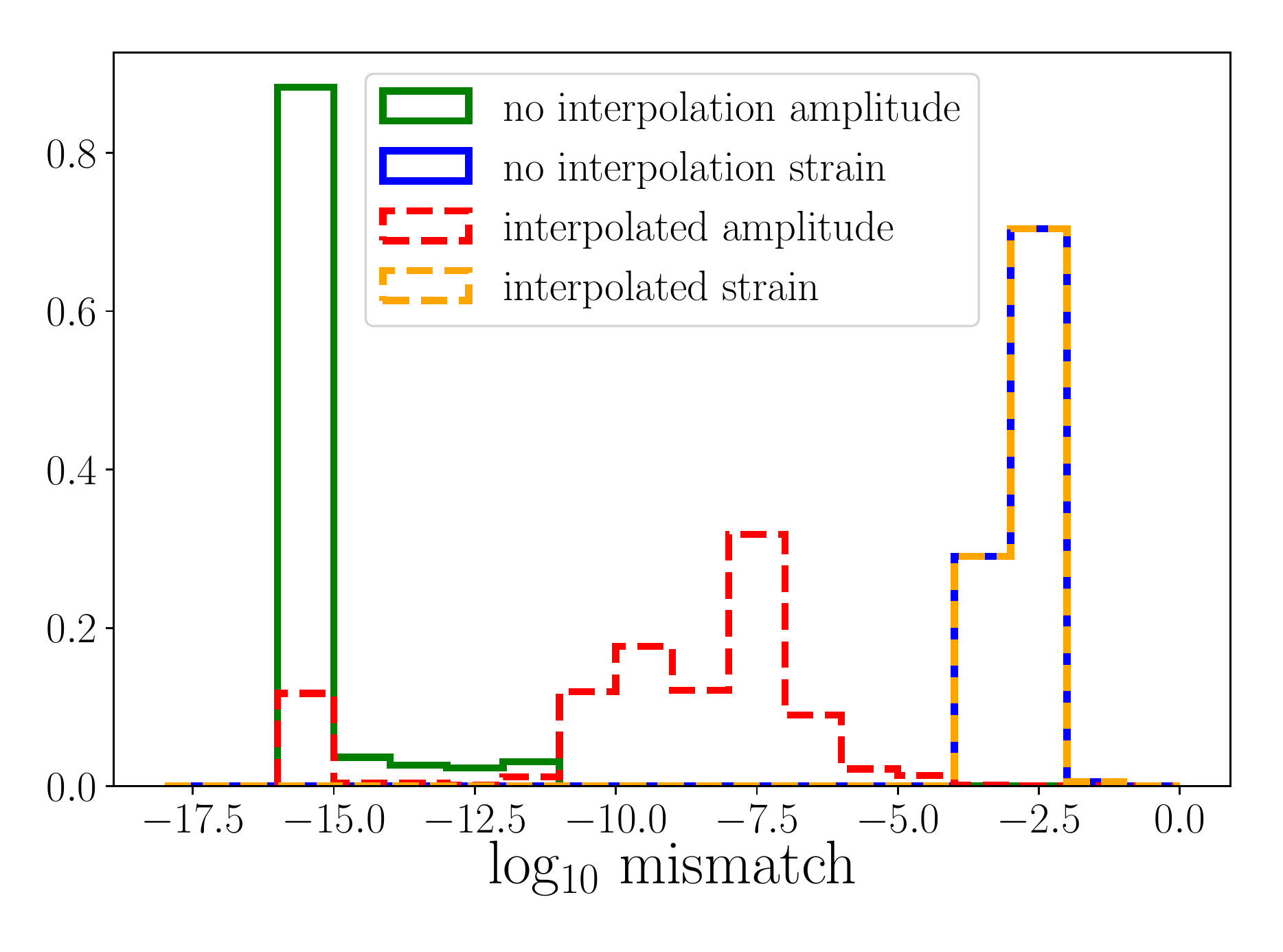}
\caption{Normalized histogram of \ac{EB} amplitude and strain mismatches 
against PhenomD in flat noise spectrum. The dashed curves are the result from 
interpolating fewer target signals; the solid line did not employ interpolation 
(cf.~Fig.~\ref{fig:match_histograms}).
}
\label{without_interp_flat}
\end{figure}

We find that the PhenomB amplitude has relatively high overlap against PhenomD 
that ranges from 90.78\% to 99.98\%. As we show in 
Fig.~\ref{without_interp_flat}, the \ac{EB} amplitude also has extremely small 
mismatches with the target signal PhenomD. Because the strain mismatches, also 
included in the figure, are orders of magnitude higher, we conclude that they 
are dominated by modelling inaccuracies in the phase.

We note that one could in principle calculate mismatches of the pure phase 
functions as well, but these numbers are less meaningful because they are not 
invariant under the physical degrees of freedom: phase and time shifts 
applied to both functions simultaneously. A geometric interpretation 
relates the overlap to `the angle' between two functions, but because the 
phases appear in the complex exponential of the strain, the relevant measure is 
the phase difference instead of their angle. 

\subsubsection{Mass scaling}
\label{subsubsec:mass_scale}

So far, we have fixed the total mass of the systems in consideration to $M = 
50M_\odot$. This choice is almost irrelevant for the actual waveform 
construction as vacuum spacetimes include the system's total mass as a simple 
scaling factor. As a result, the signal models are actually a function of the 
dimensionless product $Mf$. This degeneracy between 
total mass and frequency is broken when we need to consider physical, 
full-dimension frequencies that enter the \ac{AZDHP} noise curve. We also 
specified our lower cutoff frequency as $30 \, \rm{Hz}$. Hence, scaling the 
total mass means appropriately setting $\flow$ and $\fhigh$.

Binaries with higher total mass merge at lower frequencies. Therefore, as 
we have constructed a signal model for $M=50 M_\odot$ starting at $30\, 
\rm{Hz}$, we can use the same model also for more massive systems with the same 
$\flow$. The higher mass system then has a shorter frequency range. 

Assuming we 
have carried out the model construction for a total mass $M_1$, we can scale 
the frequency of a system with a different total mass $M_2$, but otherwise the 
same intrinsic parameters, as follows
\begin{equation}
f_2= f_1 \left(\frac{M_1}{M_2}\right).
\end{equation}
As a consequence of the Fourier transform, the strain $\tilde{h}(f)$ 
also needs to be scaled by the total mass. Putting it all together,
the strain for $M_2$ can be obtained through the following relation,
\begin{equation}
\tilde{h} (f;M_2, \eta,\chieff)= \left(\frac{M_2}{M_1}\right)^2 \; \tilde{h} 
\left(\frac{M_2 f}{M_1} ;M_1,\eta,\chieff \right).
\end{equation}

\begin{figure}
\centering
\includegraphics[width=\hsize]{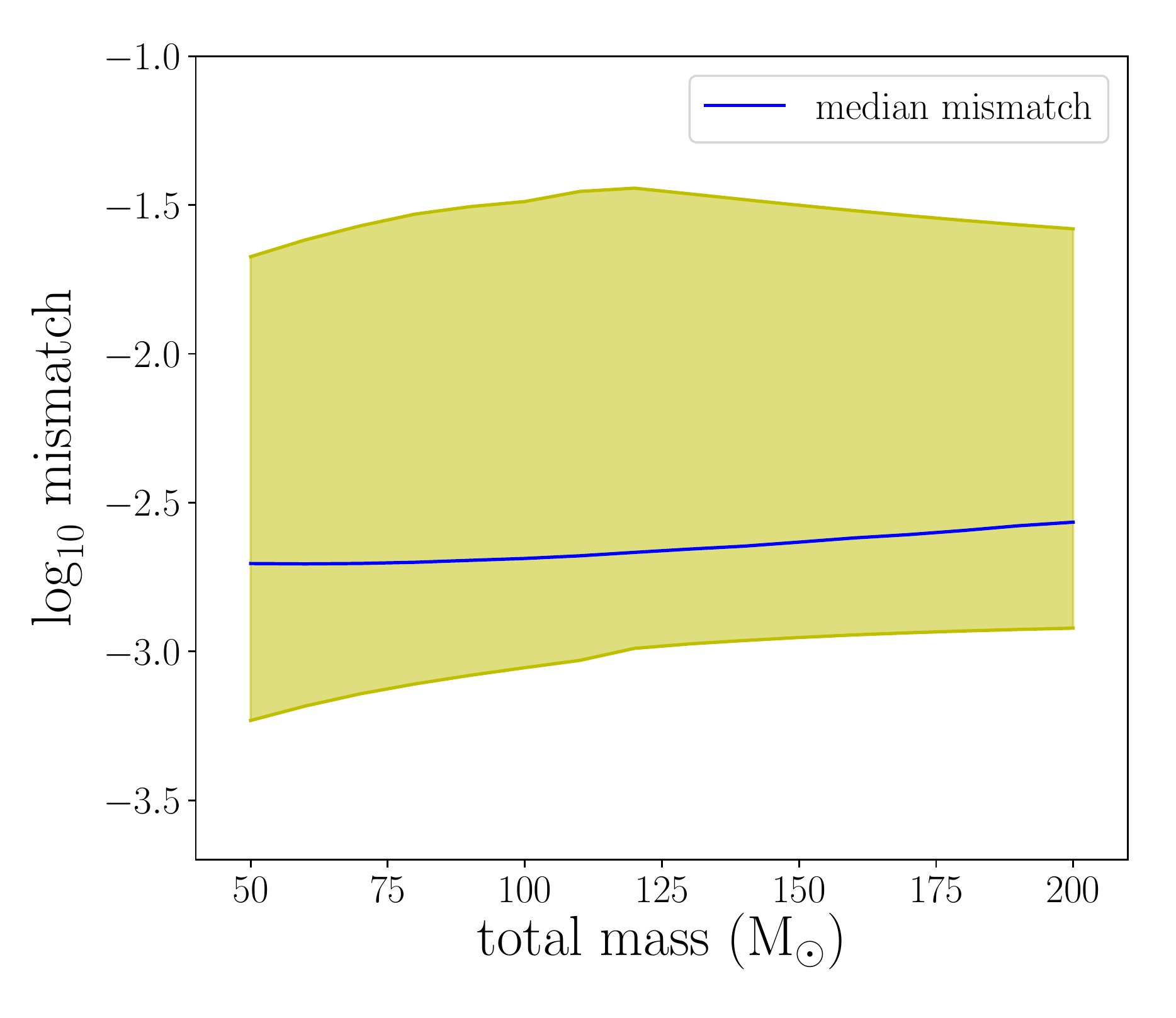}
\caption{Mismatches of EB against PhenomD for various total masses as explained in the text.
The shaded area is the range of log$_{10}$ mismatch of the respective total masses, and the blue line is its median.
This figure compares mass scaling using \ac{AZDHP} \ac{PSD} from 30 Hz to $\fhigh$ of the corresponding total mass.
}
\label{fig:mass_scaling}
\end{figure}

Without reconstructing the \ac{EB} model, we can evaluate the mismatch between 
\ac{EB} and the target model PhenomD in frequency range between $\flow$ and 
$\fhigh$ for total masses between 50 to 
200 $\Msun$. We assume the \ac{AZDHP} \ac{PSD}. The results are shown in 
Fig~\ref{fig:mass_scaling}.  
In this plot, we show that the change of mismatches are relatively small for different total masses under the \ac{AZDHP} noise spectrum.
With the same $\flow$ (30 Hz) and $\fhigh$ scaled by the total mass as explained above, higher total mass systems produce shorter waveforms. 
Since the \ac{AZDHP} noise spectrum is most sensitive in range of early hundred Hz and begin to drop gradually, the agreement between different parts of the waveforms are affected by different sensitivity ranges.
Hence the matches are not perfectly uniform for various total masses.

%\fo{Add short description of what the figure 
%tells us.}

\subsubsection{SVD iteration}
\label{subsubsec:iteration_result}

In the previous sections, we found that our method is effective in producing 
a more accurate waveform model compared to the approximate model we started 
with, PhenomB.
The mismatches of the resulting \ac{EB} family are better than PhenomB's 
mismatches against the target model, PhenomD.
This section explores a method to iterate the above steps to 
produce an even more accurate version of \ac{EB}, using the same number of 
approximate and target waveforms.

The basic idea is that we can employ the \ac{EB} model as the \emph{approximate 
waveform} of the subsequent iteration and derive a basis from $N$ \ac{EB} 
signals 
interpolated on the dense grid.
We then project the same PhenomD signals onto the new basis. We repeat this 
iterative procedure until the median does not improve significantly.

We first use the minimum number of target waveforms discussed in 
Sec.~\ref{subsubsec:min_target} and later compare the results obtained with 
more target waveforms.
We run the \ac{SVD} iteration using $12 \times 12$ PhenomD signals projected 
onto $65 \times 65$ approximate models without reducing the basis. The reported 
mismatches employ a flat noise spectrum.

\begin{figure}
\centering
\includegraphics[width=\hsize]{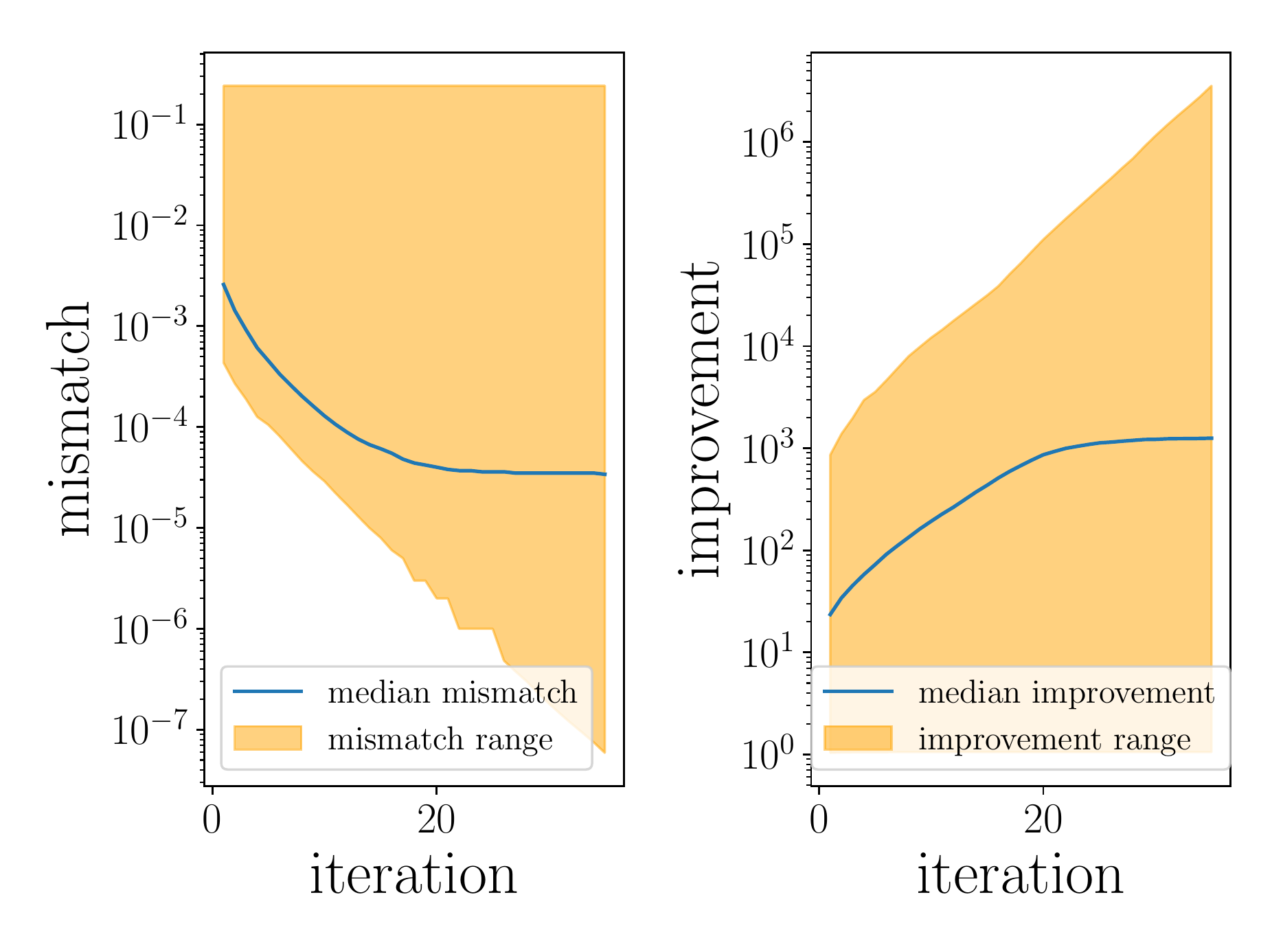}
\caption{Mismatches and improvement between EB and PhenomD after iterations. 
Left: mismatch range over iteration number shown in shaded area with 
median indicated by the blue curve.
Right: improvement range that corresponds to the same iteration is shown in shaded area, where the blue curve shows the median improvement.
%The median mismatch reduces from -2.587 to -4.463.
}
\label{mismatch_iter}
\end{figure}

The first \ac{EB} improves upon PhenomB in mismatch between 1.04 and 
860 with median of 23.5. This corresponds to $\log_{10}$ 
mismatches between $-3.36$ to $-3.57$. %after 13 minutes and 56 seconds.
We then use the \ac{EB} signals to construct a new \ac{SVD} basis and run the 
same process iteratively.
After 35 iteration the median $\log_{10}$ mismatch of \ac{EB} decreases to 
$-4.463$ while the median improvement raises to 1254.
The mismatch and improvement results are shown in Fig.~\ref{mismatch_iter}.
On a standard laptop, one iteration of this process took about 10 minutes using 
a single node (no parallelization).

For comparison, we also used PhenomD signals on a $33 \times 33$ grid and 
ran the same iterative process.
Using more target waveforms, we achieved a median mismatches
below $10^{-6}$ and an improvement of more than 1750 over PhenomB.

In conclusion, we can reduce the mismatch of \ac{EB} using an iterative 
process, but of course this will not be as effective as using more target 
waveforms.

%\begin{figure}
%\centering
%\includegraphics[width=\hsize]{../fig/improvement_12}
%\caption{The improvement grows exponentially using only the minimum of target waveforms ($12 \times 12$).
%After 35 iteration, the improvement increases from 1.035804 to 1.035804. The duration of this entire process was 7 hours and 53 minutes.
%}
%label{improvement_iter}
%\end{figure}

\subsection{Increased dimensionality: two spins}
\label{sec:doublespin}

We have shown that our method can successfully be applied to the aligned 
equal-spin case, in which both the approximate and target waveforms were varied 
across an effectively two-dimensional space of intrinsic parameters. Here
we expand the dimensionality such that the new \ac{EB} waveforms are built from 
a \emph{higher-dimensional} target model projected onto a 
\emph{lower-dimensional} approximate model.
We therefore investigate to what extent the basis can represent a greater 
parameter space than what it originated from.

Although the case we study here is not yet a practical scenario for 
actual applications, we argue that in principle
one should be able to apply this method for future projections of higher 
dimensional target models onto lower dimensional basis models.

Specifically, here we consider the case where the target waveforms PhenomD 
vary in $\eta$, $\chi_{1z}$ and $\chi_{2z}$ individually, so that $\chi_a$ [see 
Eq.~(\ref{eq:chi_a}) for its definition] does not necessarily vanish. We remind 
the reader that PhenomD is indeed sensitive to these changes, both in the 
inspiral and in predicting the ringdown signal of the remnant. In contrast, 
the approximate model PhenomB only depends on $\chieff$ and not $\chi_a$, hence 
we keep generating those signals choosing $\chieff = \chi_{1z} = 
\chi_{2z}$.
Below we discuss results and challenges of this method.

First, we generate the approximate PhenomB waveforms on the same grid of
$N = 65
\times 65$ 
points in the $\eta$-$\chieff$ parameter space that we used 
before. See Sec.~\ref{subsec:param_ranges} for details.
Second, we give ourselves $S = 33 \times 33 \times 33 = 35937$ target waveforms 
on regular 
grid $\eta$, $\chi_{1z}$ and $\chi_{2z}$.
The parameter ranges are the same as for the approximate signals, 
except that here $\chi_{1z}$, $\chi_{2z} \in [-1, 1]$ individually.
The procedure we then follow is the same as before. The \ac{SVD} basis is in 
fact unchanged compared to what we have used in previous sections, but we now 
project a much larger number of target signals onto that basis to see if we can 
accurately represent variations in a parameter that was of no relevance in the 
approximate model. 

 \begin{figure}
	\centering
\includegraphics[width=\hsize]{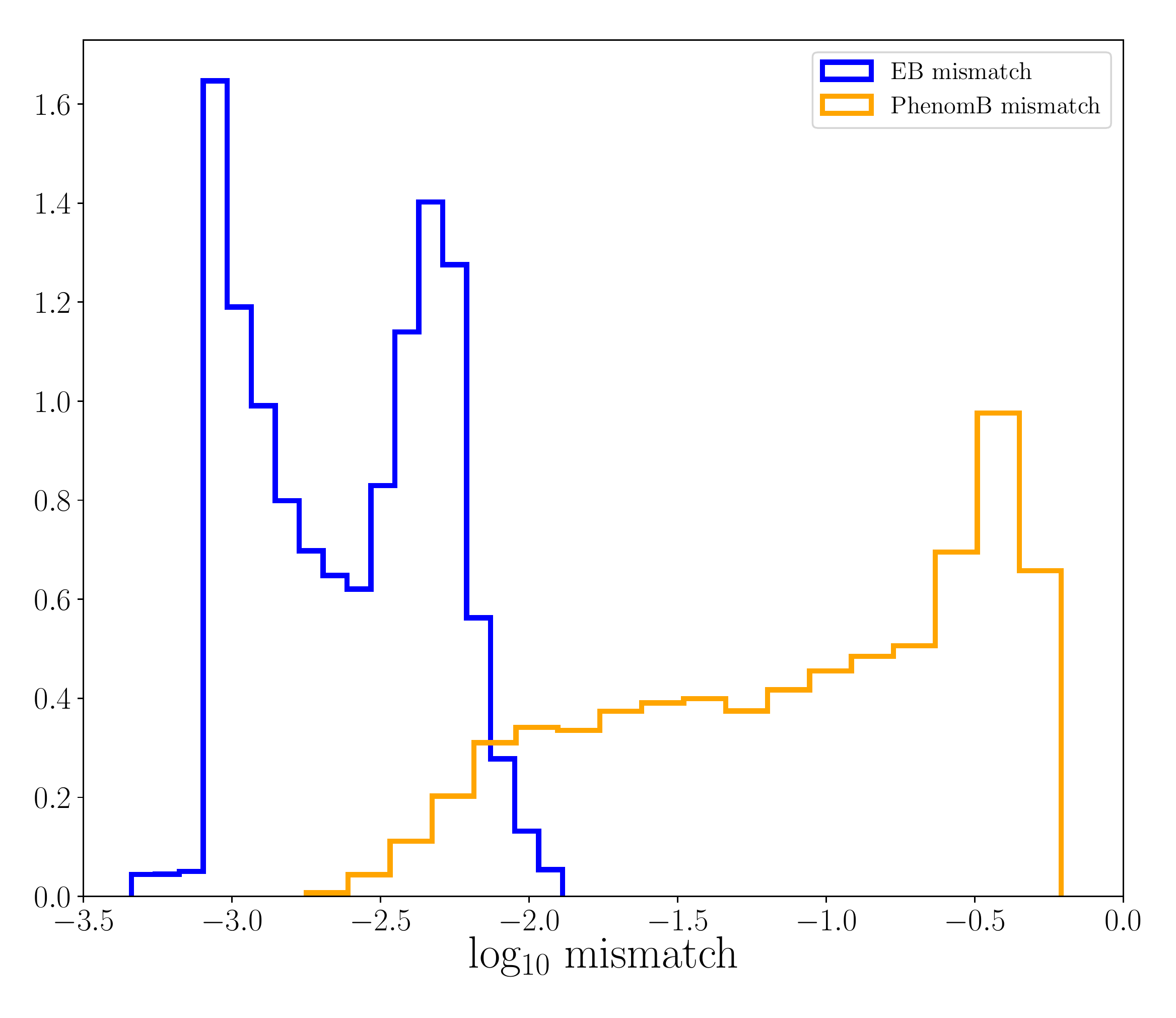}
\caption{Normalized histograms of double-spin 
\ac{EB} against PhenomD without interpolation, i.e., the \ac{EB} model was 
built from $33 \times 33 \times 33$ PhenomD waveforms projected onto an 
\ac{SVD} basis of $65 \times 65$ PhenomB signals. For comparison, the 
disagreement between PhenomB and PhenomD is also included.}
\label{double_spin_without_interp}
\end{figure}

% \subsubsection{No interpolation}
 Let us emphasize that in this study, we only analyze the errors caused by the 
projection onto a (lower-dimensional) approximate \ac{SVD} basis.
 Therefore, our comparison does not include any interpolation. Instead, we 
calculate mismatches between PhenomD and either PhenomB or \ac{EB} on all $S$ 
points of the parameter space. 
 The results are shown as histograms in 
Fig.~\ref{double_spin_without_interp}.
 The $\log_{10}$ mismatches of the \ac{EB} model range from $-1.89$ to $-3.34$ 
(which 
corresponds to matches between 0.987 to 0.999). Compared to the 
two-dimensional, equal-spin case of Sec.~\ref{subsec:single_spin}, the matches 
we find here are slightly lower. 
 This is not surprising, as here we have introduced many more PhenomD 
waveforms that we know are not accurately captured by PhenomB. 
 
 For comparison, we also show the histogram of mismatches between PhenomB and
PhenomD in the same Fig.~\ref{double_spin_without_interp}. Evidently, \ac{EB} 
achieves a much better accuracy than PhenomB, for which the $\log_{10}
$ mismatches range between 
-2.75 to -0.2 (matches between 0.380 to 0.998).
 We note that this range is similar to 
Fig.~\ref{pheB} that is restricted to the equal-spin case.

\begin{figure}
\centering
\includegraphics[width=\hsize]{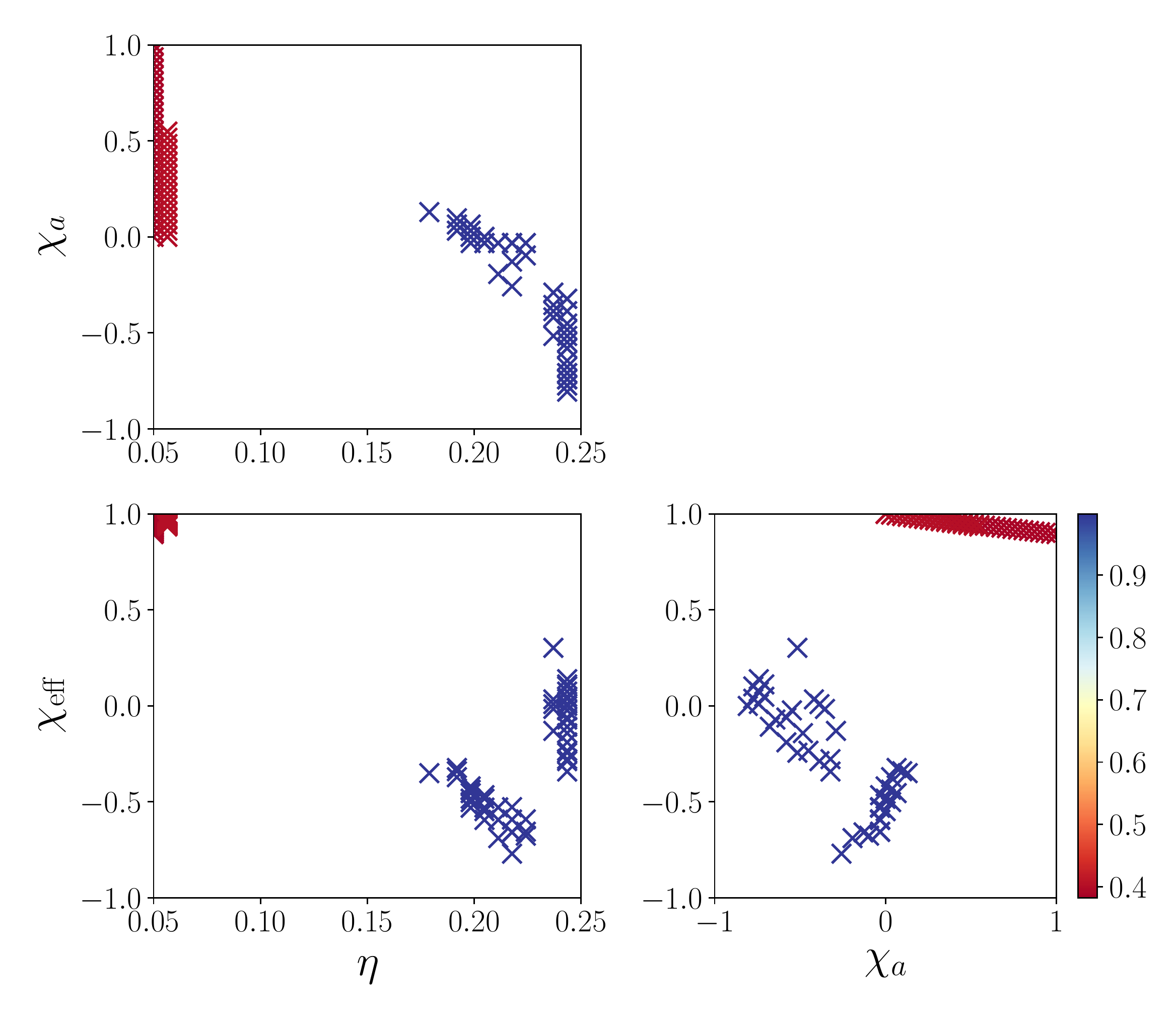}
\caption{The location of 50 lowest matches and 50 highest matches of PhenomB against PhenomD.
The colorbar shows the match.
The matches between the worst and the best waves are coloured white.}
\label{double_spin_apm_chieff}
\end{figure}

For completeness, Fig.~\ref{double_spin_apm_chieff} also illustrates the 
location of the highest mismatches of between PhenomB and PhenomD in the 
parameter space.
 In this plot, we show the location of the 50 lowest and 50 highest 
mismatches.
 The largest disagreement indeed occurs for high mass ratios and 
asymmetric spins.

  From this study, we conclude that one can in principle project a set of 
higher-dimensional 
signals onto a basis derived from a lower-dimensional model. However, 
interpolating across a high-dimensional parameter space becomes much more 
challenging, especially if a large number of bases has to be included in the 
\ac{EB} model. We leave a detailed analysis and discussion of this problem to 
future work.

\section{Conclusion and future perspectives}
\label{sec:conclusion}

The development of accurate \ac{GW} models is a crucial task to support future 
detections and the correct interpretation of \acp{GW} from merging compact 
objects.
With higher detector sensitivity in the upcoming science runs of LIGO and 
Virgo, more detections are expected, increasing the chance for an unusually 
loud, or in other ways special, observation that will require more accurate 
models than ever before.

Previous work on the development of \ac{GW} models either targeted a fairly 
restricted part of the parameter space or required substantial 
computational as well as human resources.
Here we have developed a method to dynamically update an approximate waveform 
model
in a given parameter range.
We accomplished this by projecting a set of a more 
accurate signals 
onto a larger set of a less accurate waveforms that can be 
evaluated efficiently and continuously across the parameter space. 

We worked in frequency domain and decomposed both waveform models into 
amplitude and phase that are updated separately.
Following earlier studies with a similar goal \cite{PhysRevD.87.044008}, we 
employed \ac{SVD} matrix factorization to split the approximate model's data 
into 
two unitary matrices and one diagonal matrix. We used the appropriate unitary 
matrix 
as a basis representation of the approximate model, the other two matrices are 
updated by projecting the accurate model onto that basis.
We then interpolated the projection coefficients and combined them with the 
approximate basis to obtain a new waveform 
family that we call enriched basis.
This model has a higher accuracy than the approximate model and can be 
evaluated continuously in parameter space.

In this first exploratory study, we restricted ourselves to the non-precessing 
parameter space of \acp{BBH}. We showed that 
the \ac{EB} model is considerably more faithful to its target 
model (PhenomD) than the approximate model (PhenomB) that we employed. This is 
true both for flat and \ac{AZDHP} noise spectra.
Let us highlight that especially in regions of the parameter space that were 
not accurately described by the approximate model because it had not been 
calibrated there, the improvement of \ac{EB} can be dramatic. 
This also holds if an extra physical dependence is introduced by the target 
model that was not present in the approximate model.

There are a number of procedural parameters that can be tuned in this approach 
to achieve optimal results. Among those, we tested the following.
\begin{enumerate}[(A)]
  \setlength{\itemsep}{4pt}
  \setlength{\parskip}{0pt}
 \item How many basis vectors need to be kept in the \ac{EB} model? As 
expected, we found that very few \ac{SVD} basis vectors are needed to describe 
the basic 
parameter dependence of the approximate model. However, as the success of our 
updating method relies on accurately representing effects beyond what was 
included in the approximate model, we also found that a wide parameter space 
such as 
the one tested here may require a basis of several thousand vectors. We 
expect this number to sensitively depend on the size of the parameter space and 
the accuracy of the approximate model.
 \item How many target signals are required?\\
 In the study presented here, we often used a large number of target signal to 
first test the efficacy of the basic principle. In 
Sec.~\ref{subsubsec:min_target}, we reduced that number systematically and 
analyzed the result. While the edges of the parameter space suffer increasingly 
from interpolation issues when number of target signals was reduced, we 
found that even a regular grid of $12 \times 12$ target signals showed overall 
satisfactory improvement. We note that this number is larger than the number of 
NR waveforms that were used to calibrate PhenomD~\cite{PhysRevD.93.044006, 
PhysRevD.93.044007} which is not surprising given that more physical insight 
and intuition went into the original construction while here we test an 
agnostic, fully automatic approach. We also note that we successfully tested 
uniform random placement of target waveforms instead of a regular grid, but 
designing more refined methods of placing target waveforms is an active 
research topic that can lead to a further reduction of the number of signals 
required to build 
an \ac{EB} model.
\item Can the process be iterated to achieve better results?\\
Once a fast and efficient \ac{EB} model has been built, it can and should be 
used as an approximate model for the next refinement. While this approach is 
obvious when more (or different) target waveforms become available, we also 
showed in Sec.~\ref{subsubsec:iteration_result} that such an iterative 
procedure can further improve the \ac{EB} when using the same set of target 
signals again. This might be counter-intuitive as the same target waveforms 
seem to be projected onto the same $N$-dimensional space of amplitude and phase 
functions in each iterative step. However, it turns out that performing a 
second \ac{SVD} on \ac{EB} data re-structures the basis vectors such that the 
number of irrelevant vectors with vanishingly small $\sigma$ values increases 
(i.e., the \ac{EB} is represented with fewer bases). It is these basis vectors 
that are not needed to represent the approximate model, but there are useful in 
each iterative step to slightly change the vector space toward a more faithful 
representation of the target. Further studies need to show whether such a 
procedure also introduces more irregularities and interpolation issues that 
might 
counter the gain we report here.
\end{enumerate}

Overall, the results we present here are very promising. One important
application that we work toward is actually using the best available analytical 
models as approximate signals and \ac{NR} data as the target model. In oder for 
this to be feasible, however, we need to develop additional methods in the 
immediate future. In particular, the parameter space of most interest include
precessing systems, and for those, we eventually need to deal with 
interpolating over a possibly seven-dimensional parameter space (given by two 
three-dimensional \ac{BH} spin vectors and the mass ratio). Interpolating 
a sparse set of projection coefficients (given by the available \ac{NR} 
simulations) may require much more sophisticated interpolation techniques than 
the ones we have employed here. In fact, we expect interpolation to be the 
most challenging step in more realistic applications of our procedure.

In addition, a likely scenario where our method could be extremely useful is when 
a large parameter space needs to be accessible for a signal model to be useful, 
but targeted \ac{NR} simulations only cover a reasonable small portion of that 
space. In that case, our \ac{EB} model could be updated only where new 
information is available. This can be achieved by implementing a more flexible 
interpolation approach that smoothly bridges coefficients based on the 
approximate model with information from a targeted and localized set of \ac{NR} 
data. Such a ''hybrid``\footnote{We emphasize that ''hybrids`` in \ac{GW} 
modeling commonly refer to the combination of analytical inspiral signals with 
numerical merger data. Here, however, we mean it in the sense that parameter 
space portions where only an approximate model is available could be 
smoothly combined with new information localized in certain regions of the 
parameter space.} approach would allow updating established models locally, and 
it would complement, for instance, parameter estimation methods that take 
advantage of models that can be generated for arbitrary sets of parameters \cite{PhysRevD.91.042003}
%\fo{cite lalinference} 
and alternative methods that use discrete \ac{NR} 
data sets \cite{0264-9381-34-14-144002, PhysRevD.92.023002}. %\fo{cite RapidPE etc}.

We intend to develop solutions for the above-described use cases of \ac{EB} 
in the near future. Codes will then be fully integrated in existing analysis 
suites~\cite{lalsuite} to guarantee direct impact on the analysis of \ac{GW} 
observations. We view this as an important step toward further fostering the 
integration of numerical and analytical modeling techniques in an era of 
frequent \ac{GW} observations.

\begin{acknowledgments}
The authors would like to thank to Sascha Husa, Lionel London, Harald Pfeiffer, 
and Mark Hannam for useful discussions related to this work. This work was
supported by the Max Planck Society's Independent Research Group 
Grant. Computations were carried out
on the Holodeck cluster of the Max Planck Independent Research Group ''Binary 
Merger Observations and Numerical Relativity.``

\end{acknowledgments}
\appendix
\bibliographystyle{plain}
\bibliography{vers2}

\end{document}